%% file: main.tex
\RequirePackage{fix-cm}
\documentclass[smallextended]{svjour3}       
\smartqed  
\usepackage{graphicx}
\usepackage{natbib}
\usepackage{hyperref}
\usepackage{array}
\usepackage{multicol}
\usepackage{multirow}
\usepackage{tcolorbox}
\usepackage[flushleft]{threeparttable}
\usepackage{rotating}
\usepackage{booktabs}

\begin{document}

\title{What Characteristics Make ChatGPT Effective for Software Issue Resolution? An Empirical Study of Task, Project, and Conversational Signals in GitHub Issues}

\titlerunning{Empirical Study of ChatGPT for Issue Resolution}        

\author{Ramtin Ehsani (corresponding author)         \and
        Sakshi Pathak \and
        Esteban Parra \and
        Sonia Haiduc \and
        Preetha Chatterjee
}


\institute{First Author, Second Author, Fifth Author \at
              Drexel University, Philadelphia, PA, USA \\
              \email{ramtin.ehsani@drexel.edu, sp3856@drexel.edu, preetha.chatterjee@drexel.edu}           
           \and
           Third Author \at
              Belmont University, Nashville, Tennessee, USA \\
              \email{esteban.parrarodriguez@belmont.edu}
            \and
           Fourth Author \at
              Florida State University, Tallahassee, Florida, USA \\
              \email{shaiduc@cs.fsu.edu}
}

\date{Received: date / Accepted: date}

\maketitle

\begin{abstract}
Conversational large-language models (LLMs), such as ChatGPT, are extensively used for issue resolution tasks, particularly for generating ideas to implement new features or resolve bugs. However, not all developer-LLM conversations are useful for effective issue resolution and it is still unknown what makes some of these conversations not helpful. In this paper, we analyze 686 developer-ChatGPT conversations shared within GitHub issue threads to identify characteristics that make these conversations effective for issue resolution. First, we empirically analyze the conversations and their corresponding issue threads to distinguish helpful from unhelpful conversations. \textcolor{black}{We begin by categorizing the types of tasks developers seek help with (e.g., \textit{code generation}, \textit{bug identification and fixing}, \textit{test generation}), to better understand the scenarios in which ChatGPT is most effective. Next, we examine a wide range of conversational, project, and issue-related metrics to uncover statistically significant factors associated with helpful conversations. Finally, we identify common deficiencies in unhelpful ChatGPT responses to highlight areas that could inform the design of more effective developer-facing tools.}

We found that only 62\% of the ChatGPT conversations were helpful for successful issue resolution. Among different tasks related to issue resolution, ChatGPT was most helpful in assisting with code generation, and tool/library/API recommendations, but struggled with generating code explanations. Our conversational metrics reveal that helpful conversations are shorter, more readable, and exhibit higher semantic and linguistic alignment. Our project metrics reveal that larger, more popular projects and experienced developers benefit more from ChatGPT’s assistance. \textcolor{black}{Our issue metrics indicate that ChatGPT is more effective on simpler issues characterized by limited developer activity and faster resolution times. These typically involve well-scoped technical problems such as compilation errors and tool feature requests. In contrast, it performs less effectively on complex issues that demand deep project-specific understanding, such as system-level code debugging and refactoring.} The most common deficiencies in unhelpful ChatGPT responses include incorrect information and lack of comprehensiveness. Our findings have wide implications including guiding developers on effective interaction strategies for issue resolution, informing the development of tools or frameworks to support optimal prompt design, and providing insights on fine-tuning LLMs for issue resolution tasks.
\end{abstract}

\input{1_intro}
\input{3_methods}

\input{4_results}

\input{5_threats}
\input{2_background}
\input{6_conclusion}


\bibliographystyle{spbasic}      
\bibliography{ref}   


\end{document}

%% file: 1_intro.tex
\section{Introduction}

Conversational Large Language Models (LLMs) have emerged as crucial tools for providing software development support. According to GitHub's 2023 Octoverse report, more than 92\% of developers use AI-based coding tools, including ChatGPT~\citep{noauthor_octoverse_nodate}.  
The emergence of these models has changed the way developers seek help with issue resolution, potentially replacing traditional platforms such as Stack Overflow~\citep{conversational_prog_blog}. 
Conversational LLMs are particularly valued in problem-solving tasks like issue resolution because of their ability to understand and generate human-like dialogue, thereby supporting an interactive learning experience~\citep{Ross_2023}.

Developers widely use conversational LLMs for brainstorming new features or resolving bugs \citep{hou2024large,wu2023large}. ChatGPT conversation links are shared in GitHub issue threads and pull requests to discuss potential solutions, support claims, and illustrate examples~\citep{fan2023large, tang2023empirical, hao2024empirical, 10555786}. However, these LLM responses can be of low quality or contain incorrect information~\citep{ Li2023AlwaysNA, das2024investigating, zhou2024exploringproblemscausessolutions, liu2024exploringevaluatinghallucinationsllmpowered}.
These inaccuracies are hard for developers to detect~\citep{kabir2024stack}, potentially leading to poor design decisions and software flaws, ultimately resulting in ineffective issue resolution.

There is a growing body of research on using LLMs for issue resolution with a focus on areas like bug localization~\citep{10.1145/3660773}, and program repair~\citep{SWE-bench2024, zhang2024autocoderoverautonomousprogramimprovement}. Prior research has highlighted the importance of developer conversations in successful issue resolution~\citep{CHATTERJEE2020110454, 8115620}. However, detailed analysis of conversations between developers and LLMs for issue resolution remains limited. 
To address this gap, our previous work~\citep{Ehsani25promptgaps} examined 433 developer-ChatGPT conversations in GitHub issue threads, examining how knowledge gaps in prompts and conversation styles affect issue closure.
Over 54\% of prompts exhibited knowledge gaps, including \textit{Missing Context}, \textit{Missing Specifications}, \textit{Multiple Contexts}, and \textit{Unclear Instructions}.
While we identified critical knowledge gaps in developer prompts, our analysis was limited to understanding what gaps exist and their frequency. 
We did not explore the broader interaction dynamics, such as \textit{how} developers adapt to these challenges, \textit{how} task type, project complexity, or communication patterns contribute to helpful or unhelpful outcomes, or \textit{what} strategies could mitigate their impact during multi-turn interactions. Understanding which task and project settings are more likely to lead to helpful outcomes can provide developers with actionable insights for effective issue resolution, helping them both avoid common knowledge gaps and craft better prompts.
Related research has also explored why developers engage in multi-turn conversations with ChatGPT and their motivations for sharing these conversations on GitHub \citep{hao2024empirical}.
A recent work by Das et al. revealed that ChatGPT is primarily used for ideation rather than validation, with generated code rarely being used directly to resolve issues~\citep{das2024developersengagechatgptissuetracker}. Mondal et al. identified gaps and factors that prolong these conversations \citep{mondal2024enhancing}.
While such conversations do not always lead to successful outcomes, prior studies have not investigated the specific characteristics that make some developer-LLM conversations helpful and others not, or how they contribute to effective issue resolution.

In this paper, we aim to address these gaps by analyzing a dataset of 686 issue-related conversations between developers and ChatGPT shared on GitHub issues. Our goal is to gain insights into the types of tasks related to issue resolution that ChatGPT is effective in assisting, and understanding the communication patterns and project characteristics that are associated with successful issue resolution using ChatGPT. Additionally, we identify the common deficiencies in unhelpful ChatGPT responses. By exploring the dynamics of developer-ChatGPT interactions, including how prompt characteristics (code and code-related information) and interaction strategies influence successful outcomes, we aim to uncover actionable insights to improve LLM-based issue-resolution workflows. Specifically, we investigate the following research questions:

\textit{\textbf{RQ1:} How helpful are ChatGPT conversations in successfully resolving issues?}
Using a combination of textual cues and GitHub metadata, we identify developer-ChatGPT conversations that are helpful in resolving issues and those that are not. Only 62\% of the conversations were identified as helpful. We also identify the specific tasks for which developers sought assistance during these conversations. ChatGPT was more helpful with tasks of code generation and tool/library/API recommendations, while it struggled with generating code explanations and providing up-to-date information.


\textit{\textbf{RQ2:} What are the key characteristics of issue-resolution conversations with ChatGPT? Do these characteristics vary between helpful and unhelpful conversations?}
\textcolor{black}{Using a set of conversational, project, and issue-related metrics, we investigate what differentiates helpful conversations from unhelpful ones.}
The conversational metrics are inspired by previous research on how developers ask questions on platforms like StackOverflow~\citep{10.1145/3450503, 7180105}. Our goal is to understand the challenges developers face in LLM-guided issue resolution and design metrics that could be potentially leveraged to build developer support tools.  
We found that in helpful conversations, developers provide clear, well-structured instructions and minimize topic changes within the same conversation thread.
Our project-related metrics provide insights into how project size, complexity, and popularity affect ChatGPT’s helpfulness.
We observed that larger projects and experienced developers are able to use ChatGPT more effectively for issue resolution.
\textcolor{black}{Our issue-level metrics are designed to capture the underlying complexity and nature of the issues themselves. We found that ChatGPT is more helpful on focused, lower-complexity issues that are resolved quickly and involve fewer developers. In contrast, it tends to struggle with complex, time-consuming issues that require high collaboration and deep project-specific knowledge.}

\textit{\textbf{RQ3:} What are the common deficiencies in ChatGPT's responses in unhelpful issue resolution conversations? How do they vary across different tasks?}
We found that in unhelpful issue resolution conversations, ChatGPT often produces incorrect or non-functional code, sometimes suggesting bad practices or even system-breaking solutions. Additionally, it generates unclear answers that confuse developers and hallucinates incorrect information, referencing non-existent facts or functions in various libraries.

We summarize below our key contributions in this paper:
\begin{itemize}
    \item We introduce the first manually annotated dataset of helpful and unhelpful developer-ChatGPT issue resolution conversations, consisting of 876 conversations collected from 655 projects, with 686 of the conversations from 543 projects being directly related to issues. Our code and data are made publicly available\footnote{\url{https://github.com/SOAR-Lab/llm_issue_resolution_analysis}}.
    \item \textcolor{black}{We present the first in-depth analysis of helpful vs. unhelpful developer-ChatGPT conversations for issue resolution. We identify conversational, project, and issue metrics associated with helpful conversations. We also analyze how ChatGPT performs across different issue resolution-related tasks.}
    \item We identify common deficiencies that ChatGPT exhibits in unhelpful responses related to issue resolution.
    \item We offer actionable strategies for developers to help developers obtain useful ChatGPT suggestions for issue resolution. Our findings can also inform the design of prompt optimization tools and fine-tuning LLMs for issue resolution.  
\end{itemize}


%% file: 3_methods.tex
\section{Dataset Collection and Preprocessing}

In May 2023, OpenAI introduced a feature allowing users to share their conversations with ChatGPT through dedicated links~\citep{openai2023chatgpt}. Using this feature, Xiao et al. collected a set of developer-ChatGPT conversations publicly shared on GitHub until October 2023, forming the DevGPT dataset~\citep{xiao_devgpt_2024}. 
\textcolor{black}{The size of this dataset is limited, containing only 413 conversations collected from 2023-01-01 to
2023-10-13. We expand this dataset by collecting additional data using GitHub's API endpoint, searching for occurrences of ChatGPT links in all GitHub issues. The data is collected from 2023-01-01 to 2024-04-30, encompassing the same window as DevGPT and extending it by more than six months.
We gathered all publicly available developer–ChatGPT conversation links, resulting in 876 unique conversation links from 823 GitHub issues across 655 repositories, as shown in Table \ref{tab:dataset}. Each issue thread can contain more than one link to developer-ChatGPT conversations. This dataset is not a sampled subset, but rather the result of an exhaustive crawl of all publicly shared conversation links found in GitHub issues.
While prior studies relied on the DevGPT dataset~\citep{Ehsani25promptgaps, mondal2024enhancing}, our augmented dataset more than doubles the size and offers significantly broader temporal and project-level coverage. Table~\ref{tab:dataset_comparison} provides a side-by-side comparison of the original DevGPT dataset and our extended dataset, highlighting improvements in temporal coverage, issue count, and repository diversity.}

\begin{table}[h]
\centering
\caption{\textcolor{black}{Comparison Between DevGPT and Our Extended Dataset}}
\label{tab:dataset_comparison}
\begin{tabular}{lcccc}
\toprule
\textbf{Dataset} & \textbf{Time Span} & \textbf{Issues} & \textbf{Repositories} & \textbf{Conversations} \\
\midrule
DevGPT & Jan–Oct 2023 & 388 & 300 & 413 \\
Our Dataset & Jan 2023–Apr 2024 & 823 & 655 & 876 \\
\bottomrule
\end{tabular}
\end{table}

We filtered out non-English conversations using Python's \textit{lingua-py} library \citep{stahl_pemistahllingua-py_2024}. We replaced code snippets with [CODE] and stack traces/error messages with [ERROR] inside the conversations to focus our analysis on the text rather than code and errors; the text embeds the information we are looking for, such as the description of the task, discourse style, etc. ChatGPT-generated responses structure code snippets within quote blocks, simplifying their replacement using RegEx. However, prompts are often not structured in this manner. LLMs have shown promise in automatically detecting code from unstructured documents~\citep{Oedingen_2024}. Therefore, we used \textit{GPT-4} to identify code snippets and error messages in developer prompts. To ensure the effectiveness of this method, one of the authors then conducted a manual validation for the entire dataset.

\begin{table}
\caption{Details of Our Extended Dataset}
\label{tab:dataset}
\begin{tabular}{ccccc}
\hline
\begin{tabular}[c]{@{}c@{}}\#\textbf{Unique Projects}\end{tabular} & \begin{tabular}[c]{@{}c@{}}\#\textbf{Unique Issues}\end{tabular} & \begin{tabular}[c]{@{}c@{}}\#\textbf{ChatGPT Links}\end{tabular} & \multicolumn{2}{c}{\textbf{State of Issues}} \\ \hline
655 & 823 & 876 & \begin{tabular}[c]{@{}c@{}}\#Open\\ \#Closed\end{tabular} & \begin{tabular}[c]{@{}c@{}}408\\ 415\end{tabular} \\ \hline
\end{tabular}
\end{table}

\section{Methodology}

\subsection{RQ1. How helpful are the conversations with ChatGPT in successfully resolving issues?}
We qualitatively analyzed the issue threads containing developer-ChatGPT conversations to answer RQ1, focusing on identifying the \textbf{indicators} that can help us determine the helpfulness of a ChatGPT conversation in resolving an issue and the \textbf{types of tasks} related to issue resolution presented to ChatGPT.
To determine whether a developer-ChatGPT conversation was \textit{helpful} or \textit{unhelpful}, we examined the issue thread discussions and GitHub metadata.
Specifically, we identified three types of helpfulness indicators: a) textual cues in issue threads, b) appreciation emojis in issue threads, and c) direct implementation via subsequent pull requests and/or commits. 
\textcolor{black}{These indicators emerged from inductive coding after qualitatively reviewing 100 randomly chosen issue threads. Two annotators analyzed all available metadata (e.g., comments, commits, pull requests) to identify consistent and observable signals of helpfulness. These three criteria were the only robust and repeatable markers of helpfulness across the dataset.}

\noindent
\textbf{Textual Cues:} We identified two types of textual cues.
    \noindent
    \begin{itemize}
        \item \textit{Explicit Positive Comments on ChatGPT Suggestions:} Developers explicitly praise or acknowledge the helpfulness, relevance, or accuracy of ChatGPT’s contributions inside the issue thread. Notable indicators of positive feedback included remarks such as \textit{``Chat-GPT made fast work of this one :P"} or \textit{``I should add GPT to my debug toolbox."}
        \item \textit{Using ChatGPT Suggestions For Further Exploration:} This occurred when ChatGPT's suggestions were indirectly used in two ways: 
        \textbf{(a)} Issue Discovery and Planning - information from ChatGPT helped identify issues, plan future actions and features, or explore potential solutions. For instance, comments as the following reflect proactive planning based on ChatGPT insights:  \textit{``...here’s an example of a problem that GPT-4 can easily one-shot: [Link]. In an ideal world, [name of package] could handle making the necessary edits... [Another developer] I put together a proof of concept for this: [URL]"}. 
        \textbf{(b)} Idea Generation and Brainstorming - ChatGPT guided problem-solving and brainstorming sessions. For instance, comments such as \textit{``...I’ve been studying how to get this done, these links may be useful...[Link]"} indicate ChatGPT's use for idea generation and exploration.
    \end{itemize}
        
\noindent
\textbf{Appreciation Emojis: }We identified developers' reactions to the shared ChatGPT links via emojis (e.g., hearts or likes), indicating user appreciation.

\noindent
\textbf{Direct Implementation: }We assessed helpfulness by investigating if ChatGPT’s suggestions were directly applied to resolve issues. We examined whether developers made commits or shared links to pull requests integrating the suggested code or idea, after the link was shared within the issue thread. This process involved a manual, line-by-line analysis of the pull request or commit content, cross-referenced with the specific responses provided by ChatGPT. We looked for clear alignment, including identical or near-identical code snippets, reused logic, distinctive variable or function names, and structural similarities in proposed fixes or strategies. To ensure methodological rigor, the same two authors collaboratively conducted the analysis across all candidate cases. For each instance of potential direct implementation, both authors reviewed the surrounding context (including timestamps, developer comments, and code diffs) to determine whether the changes could be attributed to the ChatGPT suggestions. Any disagreements were resolved through multiple rounds of discussion, during which both authors revisited the evidence, clarified interpretation criteria, and refined the decision logic. Final labels were assigned only after full consensus was reached.

\textcolor{black}{The three types of explicit indicators mentioned above (e.g., textual cues, appreciation emojis, direct implementation) are not mutually exclusive. A single conversation may exhibit one or multiple forms of evidence (e.g., a developer may leave a positive comment and also implement the suggestion).} We found 103 instances where code suggestions from ChatGPT lacked any of the three types of explicit indicators. In these cases, we check if the suggested code was directly integrated into the project’s codebase using a code clone detection tool.
We extracted subsequent commits made to each project's Git repository after the date when the link was shared in the thread and used the code clone detection tool NiCad Version 6.2 \citep{10.1109/ICPC.2011.26} on each commit, comparing it to the code provided by ChatGPT. To extend NiCad's support for additional languages, we added grammars based on the TXL grammar resources following their documentation\footnote{\url{https://txl.ca/txl-resources.html}}. If the tool identified a similarity between the code provided by ChatGPT and the commit made to the project, we mark the conversation as helpful. If no similarity was found, we assume that the conversation was unhelpful for the issue.
Among these 103 instances, 87 instances were written in languages supported by NiCad, either natively or with added grammars and 16 instances of unsupported languages (e.g., YAML, JSON, TypeScript) for which no grammar files were available. The distribution of programming languages was as follows: Python (49), JavaScript (15), Java (9), C++ (5), Rust (4), C (2), C\# (2), Swift (1), unsupported (16). The 16 cases in which the language was unsupported were labeled as unhelpful because no code similarity analysis could be performed and no explicit helpfulness indicators were present in the issue thread.
\textcolor{black}{For the rest, we used NiCad. The default threshold for code similarity is set to 30\%,  which is NiCad’s default setting and has been widely adopted in prior studies for detecting meaningful code clone~\citep{das2024investigating}. This threshold allows for minor edits, such as renaming variables or reformatting, while still capturing structural similarity. Given our goal of identifying cases where developers incorporated ChatGPT’s suggestions, this level of sensitivity was appropriate for our analysis.}

\textcolor{black}{All these indicators are designed to capture different forms of evidence and avoid relying on any single signal. Developers may find a response helpful without expressing explicit gratitude. To account for these variations, we combined different complementary indicators. This approach allows us to cover a wider range of helpfulness signals and reduces the likelihood of misclassification.}

\textit{Unhelpful} ChatGPT conversations are those where either the links receive negative reactions (textual or emojis) from developers, or the corresponding issue threads do not contain subsequent pull requests or commit links, and no code similarity between the suggested code and the project codebase is detected using NiCad.

\begin{table*}[]
\caption{Types of Tasks Presented to ChatGPT}
\label{tab:taxes}
\renewcommand{\arraystretch}{2} 
\large
\resizebox{\linewidth}{!}{%
\begin{tabular}{|p{3cm}|p{7cm}|p{7cm}|p{1cm}|}
\hline
\textbf{Category} & \textbf{Definition} & \textbf{\textcolor{black}{Example}} & \textbf{Freq.} \\ \hline
Code Generation and Implementation & Generation of code artifacts and scripts such as classes, functions, or modules within a project & \textit{``Create a bash script to perform server shutdown procedures"~\citep{code_generation}} & 196 \\ \hline
Tool/Library/API Recommendation & Inquiries about the usage, functionality, or integration of specific libraries, APIs, or tools & \textit{``I'm using TouchableOpacity in React, but opacity is lightened even when the user is dragging a list {[}...{]} Why is this happening?"~\citep{tool_api}} & 182 \\ \hline
\textcolor{black}{Bug Identification and Fixing} & Debugging programs, analyzing stack traces, repairing code,  and modifying code to fix bugs or errors & \textit{``Hi I'm getting these issues with fonts in CSS {[}stack trace{]}"~\citep{bug_repair}} & 120 \\ \hline
SE-Information-Seeking & Seeking information on specific or general knowledge about SE concepts or practices & \textit{``What is Hacker News, what can you tell me about the people that frequent it?"~\citep{se_information}} & 111 \\ \hline
Code Enhancement & Modifying a given code snippet to add more functionalities & \textit{``I'd like to change it {[}code{]} so that it scales values to an optionally user-specified"~\cite{code_enhancements}} & 77 \\ \hline
System Design and Architecture & Discussions about designing websites, applications, or systems, as well as recommendations for system components, package/server, and GitHub configurations & \textit{``There are 3 environments: test, development, and production. I would like to add an 'integration' environment. What would be the recommended way?"~\citep{system_design}} & 33 \\ \hline
Code Explanation & Asking for explanation of a given code inside the conversation & \textit{``I have never used html canvas before, {[}code{]}, explain it to me"~\citep{code_explain}} & 26 \\ \hline
Test Generation & Generating unit tests and test pipelines & \textit{``Now come up with a bunch of examples...Then turn those into pytest tests"~\citep{test_generation}} & 5 \\ \hline
Code Comment Generation & Generating comments for a given code & \textit{``Write docstrings for each method"~\citep{comment_generation}} & 4 \\ \hline
\end{tabular}%
}
\end{table*}

\textcolor{black}{To determine \textit{the types of SE tasks} presented to ChatGPT related to issue resolution, we analyzed the data in multiple phases. As part of this process, the annotators did not rely solely on the content of the ChatGPT conversation. They examined the broader GitHub issue thread, specifically, all comments, updates, and discussions leading up to the point where ChatGPT was invoked, to understand the context in which ChatGPT was used and to infer the task it was asked to help with. This allowed us to situate each task within the overall flow of the issue thread and capture how ChatGPT fits into real-world issue resolution practices.}

\textcolor{black}{Two authors first reviewed 100 issue threads containing ChatGPT conversation links and used open coding ~\citep{corbin_basics_2008} to identify the SE tasks.
Both annotators have 3+ years of experience in programming and qualitative analysis.
Up to two tasks were identified for each data point, as developers often had multiple requests in the same conversation. 
Analysis of the first 100 threads resulted in the identification of 12 \textit{types of tasks}. These categories started as high-level tasks such as code-related tasks or library debugging, and got merged and refined throughout the annotation process. For example, \textit{library recommendation} and \textit{library usage queries} were merged. In each of the next two phases, an additional 100 data instances (200 more data points in total) were examined independently by the two authors and discussed in iterative discussions (two in each phase) to resolve conflicts and refine the coding scheme, resulting in a final set of 9 tasks, as shown in Table \ref{tab:taxes}.
Conflicts were resolved collaboratively through consensus between the annotators.
Cohen's Kappa inter-rater agreement was calculated at each stage, reaching 0.89, a threshold supported by literature as strong for high-quality annotations~\citep{keppaArticle}. Once this level of agreement was achieved, the remaining data was divided between the two annotators for separate labeling, maintaining the consistency established earlier.}

We found that 190 conversations were unrelated to issue resolution. For instance, developers shared links with comments like \textit{“I have ChatGPT Plus. After trying to use this link, it gives me a 404"}. 
We categorized such conversations as \textit{miscellaneous} and discarded them from further analysis. We report our findings on the remaining 686 relevant conversations, shared within issue threads of 543 GitHub projects.
This remaining dataset was labeled using the nine categories described in Table \ref{tab:taxes}. Mind that the sum of the numbers on the last column in Table \ref{tab:taxes} adds up to more than 686, which is the number of issue-related links in our dataset after the removal of unrelated links, because, as mentioned before, a conversation can have more than one category assigned to it.

\subsection{RQ2. What are the key characteristics of issue-resolution conversations with ChatGPT? Do they vary across helpful and unhelpful conversations?}
\textcolor{black}{We quantitatively analyze developer-ChatGPT conversations based on three broad categories of metrics: Conversational, Project, and Issue Metrics.}
Each of these metric categories has several subcategories, as discussed in detail below.

\subsubsection{\textbf{Conversational Metrics}} 
\begin{figure}[h]
    \centering
    \includegraphics[width=1\linewidth]{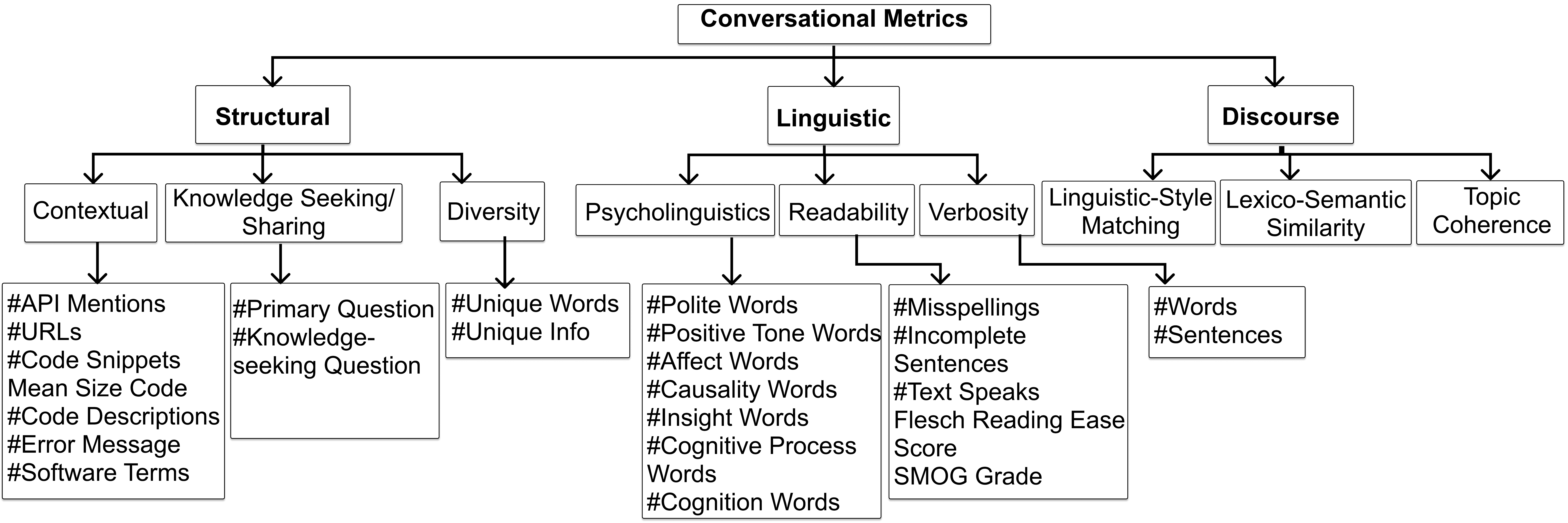}
    \caption{\textcolor{black}{Overall Structure of Conversational Metrics}}
    \label{fig:tree_diagram_metrics}
\end{figure}

We analyze the \textit{Structural}, \textit{Linguistic}, and \textit{Discourse} characteristics of the  conversations. \textcolor{black}{These metrics and their overall structure are shown in Figure \ref{fig:tree_diagram_metrics}.} 

\noindent
\underline{Structural.}
We use structural metrics to understand the types of questions developers ask, and the content they share with ChatGPT for seeking help with issue resolution. These metrics are adopted from Chatterjee et al.'s work on understanding the quality of developer chats~\citep{10.1145/3450503}. We focus on two metric categories: Knowledge-seeking and Contextual.

\textit{Knowledge-Seeking} metrics reveal the intent behind the questions developers ask ChatGPT. First, we identify the total number of \textit{\#Primary questions} in the prompts using question marks and 5W1H words~\citep{10.1145/1390334.1390415}. Next, we identify \textit{knowledge-seeking questions} by counting primary questions containing “what,” “where,” or “how”\citep{10.1145/1518701.1518819}.

\textit{Contextual} metrics provide insight into the content and characteristics of the issues with which developers seek help. We measure several SE-related information in the conversations. First, we designed regular expressions for each programming language in our dataset (e.g., Python, Java, C\#) to count \textit{\#API mentions} in the natural language. 
Next, we count \textit{\#URLs} or links using regular expressions. We also count \textit{\#Code snippets, \#Error messages}, and calculate the \textit{Mean size of code snippets} in a conversation.
To measure \textit{\#Code descriptions}, we tokenize the code identifiers from the code snippets in each conversation and count the sentences mentioning these tokens~\citep{10.1145/3450503, 7962359}. 
We calculate the \textit{\#SE terms} in the text using a pre-compiled list of morphological terms and software-related terms~\citep{7985684, dictionary_SE}.

\noindent
\underline{Linguistic.} We use linguistic metrics to assess the clarity, complexity, and overall communication effectiveness of the developer-ChatGPT conversations.  We focus on four aspects: Psycholinguistics, Diversity, Readability, and Verbosity. 

\textit{Psycholinguistic} metrics study how the human mind processes and understands language, revealing the cognitive processes behind linguistic choices. 
Using LIWC, a psycholinguistic tool widely used in software engineering studies~\citep{4228660, 10.1145/3637306, kabir2024stack, 10.1145/3544548.3581318, sandler2024linguistic}, we measure \textit{\#Polite words}, \textit{\#Positive tone words}, and \textit{\#Affect words} to understand the emotional tone of the text. Additionally, we examine the \textit{\#Causality words}, \textit{\#Insight words}, \textit{\#Cognitive process words}, and \textit{\#Cognition words} to understand the cognitive and problem-solving processes behind developers' issue resolution attempts, as exhibited in text.

\textit{Diversity} metrics quantify the unique information within each conversation to understand how frequently developers introduce new information when interacting with ChatGPT. We use \textit{\#Unique words} and \textit{\#Unique info} as indicators of diversity. We first calculate the number of unique words in the text, then determine unique information by calculating the ratio of distinct words to the total number of words\citep{10.1145/3450503}.

\textit{Readability} metrics help us assess the ease of reading information in these conversations and whether it influences the effectiveness of ChatGPT in resolving issues. 
First, using Python's pyspellchecker\citep{barrus_pyspellchecker_nodate}, we count the \textit{\#Misspelled words} in a conversation. 
Using Python's package spaCy\citep{noauthor_spacy_nodate}, we count the number of sentences without any subject or object as \textit{\#Incomplete} \citep{10.1145/3450503}.
We measure the \textit{\#Text Speaks} in a conversation by 
using a list of 50 most common text speaks and abbreviations used in chats \citep{6976134, 50_text_speaks}. Next, we measure two popular reading score metrics: \textit{Flesch Reading Ease Score} and \textit{SMOG Grade}.
We use Python's py-readability-metrics package to calculate these scores \citep{scott_smog_2023}.

\textit{Verbosity} metrics measure the length of the conversations to analyze how the amount of communication impacts the conversation outcome. We calculate the \textit{\#Words} and \textit{\#Sentences}. 

\noindent
\underline{Discourse.} We use discourse metrics to analyze how developers and ChatGPT adapt to each other's communication styles and maintain topic coherence throughout the conversation. \textcolor{black}{We begin by analyzing standard conversation statistics, including the number of words, sentences, and conversation rounds. A conversation round refers to a single exchange consisting of a user prompt followed by a ChatGPT response. We also measure the sentiment patterns using SentiStrength-SE~\citep{ISLAM2018125}, a widely used tool made for sentiment analysis in software engineering contexts~\citep{9240704}. We use a software engineering–specific sentiment analysis tool because general-purpose models often misclassify technical expressions, whereas domain-specific tools such as SentiStrength-SE are tailored to capture sentiment nuances in developer conversations~\cite{novielli2020can, imran2022data}. Overall, these metrics provide a basic understanding of conversational tone and engagement across helpful and unhelpful interactions.} 
\textcolor{black}{We capped our discourse metric-based analysis at conversation round 25, since only one unhelpful conversation in our dataset extended beyond this length.}
We focus on three key categories: Lexico-Semantic Similarity, Linguistic-Style Matching, and Topic Coherence.  

\textit{Lexico-Semantic Similarity} measures the semantic similarity between developer prompts and ChatGPT's answers throughout the conversation. We track how this score changes at each step to assess if the conversation stays semantically aligned. Using word embeddings from a pre-trained model optimized for semantic search \citep{all-minilm-l6-v2_2024}, we compute the cosine similarity between the prompts and responses~\citep{10.1145/3411764.3445645}.

\textit{Linguistic-Style Matching (LSM)} evaluates the similarity of linguistic style between developers and ChatGPT. We examine if developers and ChatGPT adjust their word choice and syntax to match each other at each step. We analyze each developer prompt with its corresponding ChatGPT answer and each prompt with the previous ChatGPT answer to track LSM changes. The score is calculated using 8 non-topical categories: articles, conjunctions, impersonal pronouns, personal pronouns, negations, prepositions, adverbs, and auxiliary verbs \citep{10.1145/1963405.1963509}. A higher score indicates more LSM \citep{10.3389/fpsyg.2022.949968, boyd_liwc}.


\textit{Topic Coherence} refers to how the discourse evolves over several turns, maintaining focus on the same topic or diverging into others. \citep{vakulenko2018measuring}. To evaluate coherence in developer-ChatGPT conversations, we used topic modeling with Python's Gensim library~\citep{10.1145/2684822.2685324}. At each step, we analyzed the last two exchanges using Latent Dirichlet Allocation (LDA) to calculate semantic similarity between high-scoring words in the topics~\citep{10.5555/944919.944937}. For LDA, we used 1 topic and set the number of passes to 10, as this configuration showed to yield stable topic distributions for our task. The decision to focus on the last two exchanges (steps) was deliberate. We wanted to capture any sudden shifts in the topic that might occur between consecutive prompt-answer pairs, as these shifts are key indicators of whether the conversation is drifting off-topic. Analyzing the entire conversation would obscure such shifts by aggregating too much data, potentially masking important transitions. A higher coherence score represents participants' focus on one technical topic and less drift to off-topic discussions.
\subsubsection{\textbf{Project Metrics}} We analyze the \textit{Repository Parameters} and \textit{Developer Experience} characteristics of the projects involved in the conversations.

\noindent
\underline{Repository Parameters.} We examine if the size, maturity, and popularity of the GitHub projects influence the perception of ChatGPT's helpfulness in issue resolution. To account for project size and maturity, we measure: \textit{\#Stars}, \textit{\#Contributors}, \textit{\#Forks}, \textit{\#Files}, and \textit{\#Lines of Code}. As indicators of project popularity we measure: \textit{\#Stars}, \textit{\#Contributors}, and \textit{\#Forks}. We extracted these metrics using GitHub's API, making specific calls for each project in our dataset. To count \textit{\#Files} and \textit{\#Lines of Code}, we cloned the repositories and ran the tool \textit{cloc}~\citep{adanial_cloc}. 

\noindent
\underline{Developer Experience.} 
We identify the developers who first shared the links to the ChatGPT conversations in the issue threads and then examine if their level of experience influences the success of the conversations. We assume that the developer who first shares the link is the one who had the conversation with ChatGPT. Using GitHub's API, we extract the number of public repositories created by the developer (\textit{\#Public Repositories}), the number of followers the developer has (\textit{\#Followers}), the time difference between when the developer shared the link and when they first created their account (\textit{Account Age}), and the total number of push commits and issue thread/pull request activities the developer has had on GitHub in the past year (\textit{\#Contributions}) from their public accounts.

\textcolor{black}{\subsubsection{\textbf{Issue Metrics}}
For each conversation, we analyze the corresponding GitHub issue threads to investigate two key characteristics: \textit{Issue Difficulty} and \textit{Issue Type}. These characteristics help us examine whether certain types of issues or levels of complexity are associated with helpful or unhelpful ChatGPT interactions.}

\noindent
\textcolor{black}{
\underline{Issue Difficulty.}
We assess whether ChatGPT is more helpful when assisting with simpler or more complex issues by extracting several features from the issue threads as follows: i) Total number of comments exchanged within the thread (\textit{\#Comments}), ii) the total word count across all comments (\textit{Length of Discussion}), iii) time elapsed between the first and last comment (\textit{\#Discussion Hours}), iv) the number of distinct developers participating in the discussion (\textit{\#Unique Developers}), v) if the issues are resolved and patched, time taken to close the issue (\textit{Resolution time}), vi) the size of the patch based on the number of lines added and removed inside code (\textit{Patch Size})~\citep{ren2020understandingnaturesystemrelatedissues, 6976094, 10992485, 10.1145/3385032.3385052}. Together, these metrics provide a multifaceted view of issue complexity and effort, helping us assess how task difficulty may influence ChatGPT’s helpfulness.}

\noindent
\textcolor{black}{
\underline{Issue Type.}
To explore the types of issues developers get help with from ChatGPT, we classify issues into thematic categories using unsupervised clustering. We first extract the title and description of each issue and transform them into TF-IDF vectors~\citep{app13158788}. To reduce dimensionality and capture latent semantic patterns, we apply Truncated SVD (LSA)~\citep{Al_Msie_deen_2024, franca_gpts_2025}.}

\begin{figure}[h]
    \centering
    \includegraphics[width=1\linewidth]{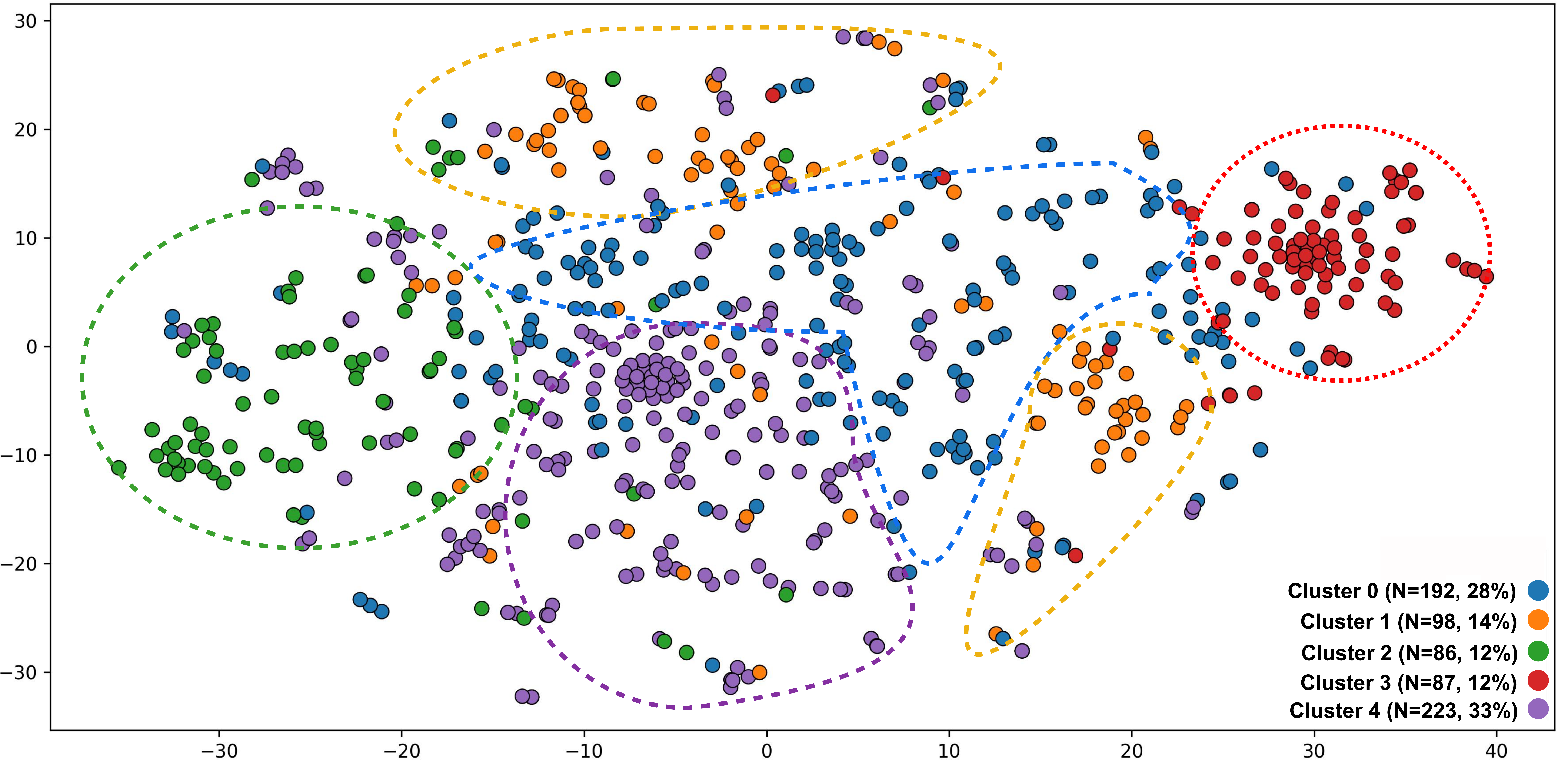}
    \caption{\textcolor{black}{Final Clusters using K-means}}
    \label{fig:clusters_2}
\end{figure}

\textcolor{black}{
Initially, we experimented with HDBSCAN, a density-based clustering algorithm that does not require specifying the number of clusters. While HDBSCAN is generally effective for discovering arbitrarily shaped clusters and filtering noise, we found that it performed poorly on our data due to its sparse and high-dimensional nature. The resulting clusters were overly generalized and noisy, often grouping semantically unrelated issues together or labeling most of the data as outliers. This made it difficult to interpret and assign meaningful categories.
Therefore, we opted for K-Means, which is better suited for well-separated, high-dimensional data like TF-IDF representations. K-Means provides clearer and more consistent partitions, allowing us to interpret the resulting clusters more reliably. Using the elbow method~\citep{Umargono2020}, we determined that k = 8 offered the best trade-off between compactness and separation. In the initial clustering results, the t-SNE (t-distributed Stochastic Neighbor Embedding) projection revealed meaningful groupings but also some notable overlaps between clusters. We observed substantial thematic overlap between clusters 0, 2, 4, and 7, all of which contained variants of improvement-oriented issues (e.g., API feature requests, documentation suggestions, minor usability fixes). To improve interpretability, we merged these four clusters into a single group (cluster 0), resulting in a more coherent and distinct set of themes. The refined clusters, as shown in Figure~\ref{fig:clusters_2}, consist of a total of 5 clusters, revealing better visual separation and alignment with underlying issue types. }


\textcolor{black}{
To assign interpretable labels to each cluster, we employ a semi-automated labeling approach. LLMs have shown strong performance in annotating and classifying software engineering data, making them a promising tool for this task~\citep{ahmed2025llmsreplacemanualannotation, 10.1145/3643991.3644903, tan2024largelanguagemodelsdata}. For each cluster, we sample 20 random issues and pass them to GPT-4o with a prompt asking for a short, specific label that captures the type of issue.
With cluster sizes ranging from 86 to 223, a random sample of 20 issues yields a margin of error between approximately ±19\% and ±21\% at a 95\% confidence level (assuming maximum variability at p=0.5). This provides a reasonable balance between coverage and feasibility for high-level thematic labeling.
This sampling strategy also follows prior work that found 10–20 examples (or even top keywords) per cluster is sufficient for consistent and interpretable labeling, especially when combined with human validation~\citep{eklund-forsman-2022-topic, Miller_2025, khandelwal2025usingllmbasedapproachesenhance}. Therefore, we also manually verified and refined the generated labels by reviewing the cluster contents and ensuring consistency with the developer's intent.}

\subsection{RQ3. What are the common deficiencies in ChatGPT's responses in unhelpful issue-resolution conversations? How do they vary across different tasks?}

Recent studies have identified several deficiencies in ChatGPT's responses to software engineering tasks. For instance, they tend to be more verbose and comprehensive than human answers, but often include incorrect information and inconsistencies \citep{kabir2024stack, Li2023AlwaysNA}. Other issues are hallucinations, conflicting intents, knowledge gaps, and context deviations \citep{liu2024exploringevaluatinghallucinationsllmpowered}.
Building on these prior works, 
we examine the issue thread discussions of unhelpful developer-ChatGPT conversations, looking for reasons why developers found the conversations unhelpful. 
\textcolor{black}{Specifically, we qualitatively analyze the negative signals in the issue threads that follow shared ChatGPT conversation links. Comments such as \textit{``ChatGPT is wrong"}~\citep{chatgpt_is_wrong} or \textit{``not sure how trustworthy that info is"}~\citep{chatgpt_trust} are examples of such indicators in our dataset. Similar to RQ1, to derive a taxonomy of deficiencies, we applied iterative open coding on a random sample of 100 unhelpful instances. The same two annotators independently reviewed each data point and assigned codes based on observed issues in the ChatGPT responses. After each round, they discussed disagreements, refined the code definitions, and repeated the process. This continued until they reached a stable set of five deficiency categories and achieved a Cohen’s Kappa inter-rater agreement of 0.87, indicating strong reliability in the coding process~\cite{keppaArticle}.
To measure how the deficiencies vary across tasks, we use the taxonomy used in Table \ref{tab:taxes} and calculate the frequency of deficiencies in each category.}

%% file: 4_results.tex
\section{Results}
 
\subsection{RQ1. How helpful are the conversations with ChatGPT in successfully resolving issues?}\label{sec:rq1_1}


\begin{figure}[ht]
    \centering
    \includegraphics[width=0.95\linewidth]{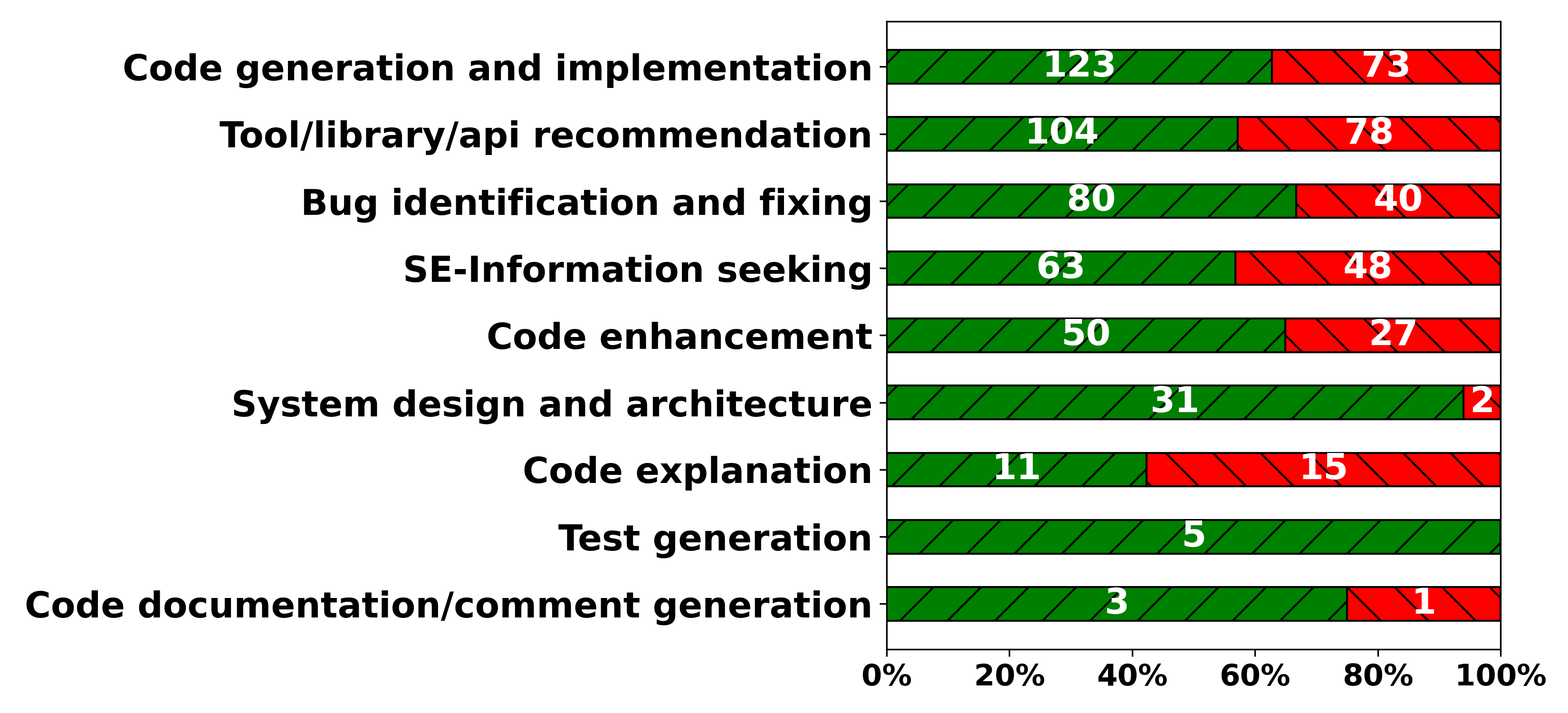}
    \vspace{-0.2cm}
    \caption{Number and Percentage of Helpful and Unhelpful Conversations for Each Task. \textcolor{teal}{\textbf{Green}} Accounts for Helpful and \textcolor{red}{\textbf{Red}} Accounts for Unhelpful.}
    \label{fig:proportion}
    \vspace{-0.2cm}
\end{figure}

We identified nine types of SE tasks for which developers sought help from ChatGPT (see Table \ref{tab:taxes}). The most frequent tasks include \textit{Code Generation and Implementation} (196 instances), \textit{Tool/Library/API Recommendation} (182 instances), and \textit{Bug Identification and Fixing} (120 instances). 
Given that issue threads typically revolve around code and debugging, and that LLMs are known to excel in these tasks~\citep{hou2024large}, it is expected that such tasks would be more frequent in our dataset. 
The least observed tasks were \textit{Test Generation} and \textit{Code Comment Generation}, occurring only in 5 and 4 instances, respectively. 

\begin{figure}[h]
    \centering
    \includegraphics[width=0.95\linewidth]{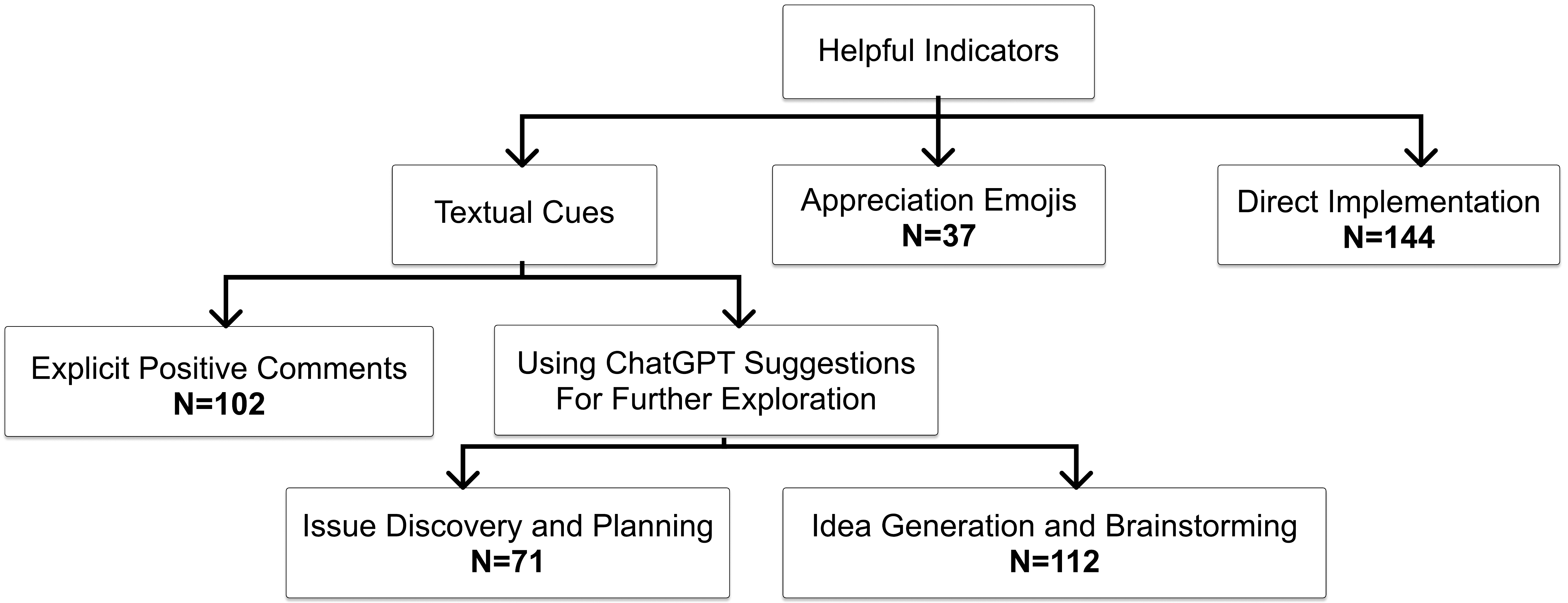}
    \caption{Helpful Indicators for Developer-ChatGPT Conversations and their Distribution}
    \label{fig:tree_diagram}
\end{figure}

Only 428 out of 686 developer-ChatGPT conversations (62\%) were found to be helpful to developers, while 258 were not. This is unexpected given that shared links imply the poster believed the links would be beneficial. The helpful indicators and their distribution are shown in Figure \ref{fig:tree_diagram}. 
The most common indicators include: (1) Direct Implementation ($n=144$), and (2) Textual Clues, a broader category that includes \textit{Idea Generation and Brainstorming} ($n=112$) and \textit{Explicit Positive Comments} ($n=102$).
In the case of 103 code-related conversations with ChatGPT, where they lacked clear indicators to determine whether the conversation was beneficial to the issue thread, after running NiCad, only in two instances significant similarities (88\% and 100\%) were found between the ChatGPT-provided code and the commits.

The perceived efficacy of ChatGPT conversations (percentage of helpful vs. unhelpful) across different issue resolution tasks is shown in Figure \ref{fig:proportion}. 
ChatGPT was most effective in addressing \textit{Code Generation and Implementation}, with 123 conversations deemed helpful (62\%) and 73 unhelpful. The next most effective tasks were \textit{Tool/Library/API Recommendation} (57\% helpful) and \textit{Bug Identification and Fixing} (66\% helpful). 
Developers appreciated the depth of insight provided by ChatGPT, with one stating, \textit{``Wow. That is scary and also very informative"}~\citep{this_is_scary} after using ChatGPT to debug DNS null entries in their application. They also praised the adaptability of ChatGPT, with one noting \textit{``[ChatGPT] do fairly well going from one syntax to the other"}~\citep{fairly_well} after using ChatGPT to translate an Apache config to Nginx. The diagnostic capabilities of ChatGPT in detecting errors were also noticed by developers, with one saying \textit{``I'm definitely adding ChatGPT to my debug toolbox"}~\citep{toolbox} after resolving a series of errors and successfully getting their site running locally.
\textcolor{black}{ChatGPT also performed well on \textit{System Design and Architecture} tasks, with 31 out of 33 conversations (94\%) labeled as helpful. While this may seem surprising given the high-level system knowledge typically required for such tasks, our analysis revealed that most queries in this category were exploratory in nature. Developers sought brainstorming support, general design strategies, or best practices for implementing architectural components. In these cases, developers often provided sufficient high-level context, enabling ChatGPT to suggest flexible solution paths and design options. This ideation support was often valued by developers, even if the model lacked access to the full system context. However, we note that the relatively small sample size for this task type (n=33) may also contribute to the high success rate observed.}

ChatGPT struggled the most with \textit{Code Explanation}, the only task where the number of unhelpful conversations (n=15) was more than the helpful (n=11). 
In many unhelpful conversations, code snippets were presented to ChatGPT without sufficient context about the codebase or the issue at hand. This affected the accuracy of the answers provided by ChatGPT. 
ChatGPT also struggled with \textit{SE-Information Seeking}, where developers used ChatGPT to gather specific information related to an issue. For example, developers asked ChatGPT about WebAssembly's memory usage in Chrome. However, conflicting information from other sources led them to question ChatGPT's reliability~\citep{information_assembly}. \textcolor{black}{Based on such patterns in our dataset, using ChatGPT as a search engine for issue resolution may not be recommended, as the model’s inaccuracies, especially in factual queries, can be difficult to detect. While there are instances where ChatGPT was helpful in this role, the risk of subtle or undetected misinformation makes this usage mode less reliable for issue resolution.}

\textcolor{black}{Our results also indicate that \textit{Test Generation} and \textit{Code Documentation} were relatively uncommon tasks for which developers engaged ChatGPT in our dataset. While these areas have been well-studied in the broader research community and supported by various LLM-based tools~\citep{10329992}, we observed few instances of developers using ChatGPT conversationally for these purposes during issue resolution. In the limited cases where ChatGPT was applied to these tasks, developers generally found its responses helpful, suggesting that there may be potential in leveraging ChatGPT for these use cases in practice.}

\begin{tcolorbox}[]
428 out of 686 developer-ChatGPT conversations were found helpful in issue resolution (62\%). ChatGPT was most effective in tasks like \textit{Code Generation and Implementation}, \textit{Tool/Library/API Recommendation}, and \textit{Bug Identification and Fixing}. However, it struggled with \textit{Code Explanation} and \textit{SE Information-Seeking}. 
\end{tcolorbox}

\subsection{RQ2. What are the key characteristics of issue-resolution conversations with ChatGPT? Do they vary across helpful and unhelpful conversations?}
First, we discuss the differences in Conversational Metrics (Structural, Linguistic, Discourse) between helpful and unhelpful conversations. Next, we compare the Project Metrics (Repository, Developer), \textcolor{black}{and finally, we look into Issue Metrics (Difficulty, Type). Our analysis identifies features that commonly appear in helpful conversations to help characterize patterns in how successful conversations tend to unfold.}

\begin{table*}[]
\renewcommand{\arraystretch}{1}
\Huge
\caption{Relative Differences between Helpful and Unhelpful Developer-ChatGPT Conversations for Conversational Metrics in Developer Prompts and ChatGPT Answers. \textbf{Positive} Numbers Indicate Higher Frequencies of Metrics, and \textbf{Negative} Numbers Indicate Lower Frequencies in Helpful Conversations.}
\label{tab:rq2_1}
\resizebox{\columnwidth}{!}{%
\begin{tabular}{|l|l|ccc|ccc|}
\hline
\multirow{2}{*}{\textbf{Metrics}} & \multirow{2}{*}{\textbf{Sub-categories}} & \multicolumn{3}{c|}{\textbf{Developers}} & \multicolumn{3}{c|}{\textbf{ChatGPT}} \\ \cline{3-8} 
 &  & \multicolumn{1}{c|}{\textbf{\begin{tabular}[c]{@{}c@{}}Corrected\\ P-Value\end{tabular}}} & \multicolumn{1}{c|}{\textbf{Rel. Diff.}} & \textbf{\begin{tabular}[c]{@{}c@{}}Range\\ (Min\textless{}Mdn\textless{}Max)\end{tabular}} & \multicolumn{1}{c|}{\textbf{\begin{tabular}[c]{@{}c@{}}Corrected\\ P-Value\end{tabular}}} & \multicolumn{1}{c|}{\textbf{Rel. Diff.}} & \textbf{\begin{tabular}[c]{@{}c@{}}Range\\ (Min\textless{}Mdn\textless{}Max)\end{tabular}} \\ \hline
\begin{tabular}[c]{@{}l@{}}\textbf{Structural:}\\ Knowledge\\ Seeking/\\ Sharing\end{tabular} & \begin{tabular}[c]{@{}l@{}}\#Primary Question\\ \#Knowledge-seeking Question\end{tabular} & \multicolumn{1}{c|}{\begin{tabular}[c]{@{}c@{}}0.27\\ 0.68\end{tabular}} & \multicolumn{1}{c|}{\begin{tabular}[c]{@{}c@{}}0.25\\ -0.02\end{tabular}} & \begin{tabular}[c]{@{}c@{}}0\textless{}1\textless{}68\\ 0\textless{}0\textless{}13\end{tabular} & \multicolumn{1}{c|}{\begin{tabular}[c]{@{}c@{}}-\\ -\end{tabular}} & \multicolumn{1}{c|}{\begin{tabular}[c]{@{}c@{}}-\\ -\end{tabular}} & \begin{tabular}[c]{@{}c@{}}-\\ -\end{tabular} \\ \hline
\begin{tabular}[c]{@{}l@{}}\textbf{Structural:}\\ Contextual\end{tabular} & \begin{tabular}[c]{@{}l@{}}\#API Mentions\\ \#URLs\\ \#Code Snippets\\ Mean Size Code Snippets\\ \#Code Descriptions\\ \#Error Message\\ \#Software-specific Terms\end{tabular} & \multicolumn{1}{c|}{\begin{tabular}[c]{@{}c@{}}0.02\\ 0.68\\ 0.02\\ 0.02\\ 0.05\\ 0.40\\ 0.04\end{tabular}} & \multicolumn{1}{c|}{\begin{tabular}[c]{@{}c@{}}\textbf{-0.39**}\\ 0.16\\ \textbf{0.11**}\\ \textbf{-0.13**}\\ \textbf{-0.21*}\\ 0.07\\ \textbf{-0.17**}\end{tabular}} & \begin{tabular}[c]{@{}c@{}}0\textless{}0\textless{}126\\ 0\textless{}0\textless{}39\\ 0\textless{}1\textless{}130\\ 0\textless{}18.625\textless{}17713\\ 0\textless{}0\textless{}1242\\ 0\textless{}0\textless{}18\\ 0\textless{}7.5\textless{}463\end{tabular} & \multicolumn{1}{c|}{\begin{tabular}[c]{@{}c@{}}0.99\\ 0.88\\ 0.23\\ 0.45\\ 0.31\\ -\\ 0.75\end{tabular}} & \multicolumn{1}{c|}{\begin{tabular}[c]{@{}c@{}}0.14\\ -0.17\\ \textbf{0.28*}\\ 0.04\\ 0.05\\ -\\ 0.001\end{tabular}} & \begin{tabular}[c]{@{}c@{}}0\textless{}2\textless{}61\\ 0\textless{}0\textless{}9\\ 0\textless{}10\textless{}611\\ 0\textless{}56\textless{}6294\\ 0\textless{}28\textless{}3917\\ -\\ 0\textless{}44\textless{}501\end{tabular} \\ \hline
\begin{tabular}[c]{@{}l@{}}\textbf{Linguistic:}\\ Psycholing-\\ uistics\end{tabular} & \begin{tabular}[c]{@{}l@{}}\#Polite Words\\ \#Positive Tone Words\\ \#Affect Words\\ \#Causality Words\\ \#Insight Words\\ \#Cognitive Process Words\\ \#Cognition Words\end{tabular} & \multicolumn{1}{c|}{\begin{tabular}[c]{@{}c@{}}0.04\\ 0.05\\ 0.04\\ 0.02\\ 0.05\\ 0.03\\ 0.02\end{tabular}} & \multicolumn{1}{c|}{\textbf{\begin{tabular}[c]{@{}c@{}}0.73**\\ 0.48*\\ 0.32**\\ -0.23**\\ -0.24*\\ -0.10**\\ -0.09**\end{tabular}}} & \begin{tabular}[c]{@{}c@{}}0\textless{}0\textless{}33.33\\ 0\textless{}0\textless{}33.33\\ 0\textless{}0.545\textless{}33.33\\ 0\textless{}2.975\textless{}30\\ 0\textless{}2.11\textless{}25.0\\ 0\textless{}12.48\textless{}37.5\\ 0\textless{}12.94\textless{}37.5\end{tabular} & \multicolumn{1}{c|}{\begin{tabular}[c]{@{}c@{}}0.74\\ 0.99\\ 0.94\\ 0.50\\ 0.98\\ 0.99\\ 0.65\end{tabular}} & \multicolumn{1}{c|}{\begin{tabular}[c]{@{}c@{}}-0.03\\ 0.05\\ 0.01\\ -0.06\\ -0.07\\ -0.03\\ -0.03\end{tabular}} & \begin{tabular}[c]{@{}c@{}}0\textless{}0\textless{}7.43\\ 0\textless{}0\textless{}2.5\\ 0\textless{}1.26\textless{}11.11\\ 0\textless{}3.16\textless{}9.69\\ 0\textless{}2.09\textless{}11.54\\ 0\textless{}14.15\textless{}25.42\\ 0\textless{}14.42\textless{}25.81\end{tabular} \\ \hline
\begin{tabular}[c]{@{}l@{}}\textbf{Linguistic:}\\ Diversity\end{tabular} & \begin{tabular}[c]{@{}l@{}}\#Unique Words\\ \#Unique Info\end{tabular} & \multicolumn{1}{c|}{\begin{tabular}[c]{@{}c@{}}0.06\\ 0.34\end{tabular}} & \multicolumn{1}{c|}{\begin{tabular}[c]{@{}c@{}}\textbf{-0.22*}\\ -0.11\end{tabular}} & \begin{tabular}[c]{@{}c@{}}1\textless{}36.5\textless{}2233\\ 1\textless{}11\textless{}212\end{tabular} & \multicolumn{1}{c|}{\begin{tabular}[c]{@{}c@{}}0.99\\ 0.97\end{tabular}} & \multicolumn{1}{c|}{\begin{tabular}[c]{@{}c@{}}-0.03\\ 0.03\end{tabular}} & \begin{tabular}[c]{@{}c@{}}1\textless{}158\textless{}2016\\ 1\textless{}6.66\textless{}64.33\end{tabular} \\ \hline
\begin{tabular}[c]{@{}l@{}}\textbf{Linguistic:}\\ Readability\end{tabular} & \begin{tabular}[c]{@{}l@{}}\#Misspellings\\ \#Incomplete Sentences\\ \#Text Speaks\\ Flesch Reading Ease Score\\ SMOG Grade\end{tabular} & \multicolumn{1}{c|}{\begin{tabular}[c]{@{}c@{}}0.79\\ 0.06\\ 0.88\\ 0.84\\ 0.04\end{tabular}} & \multicolumn{1}{c|}{\begin{tabular}[c]{@{}c@{}}-0.23\\ \textbf{-0.63*}\\ 0.50\\ 0.03\\ \textbf{0.20**}\end{tabular}} & \begin{tabular}[c]{@{}c@{}}0\textless{}1\textless{}145\\ 0\textless{}0\textless{}240\\ 0\textless{}0\textless{}6\\ -790\textless{}68.77\textless{}121\\ 0\textless{}7.45\textless{}59.9\end{tabular} & \multicolumn{1}{c|}{\begin{tabular}[c]{@{}c@{}}-\\ -\\ 0.99\\ 0.91\\ 0.89\end{tabular}} & \multicolumn{1}{c|}{\begin{tabular}[c]{@{}c@{}}-\\ -\\ 0.11\\ 0.02\\ -0.02\end{tabular}} & \begin{tabular}[c]{@{}c@{}}-\\ -\\ 0\textless{}0\textless{}8\\ -5\textless{}56.93\textless{}121\\ 0\textless{}12\textless{}22.5\end{tabular} \\ \hline
\begin{tabular}[c]{@{}l@{}}\textbf{Linguistic:}\\ Verbosity\end{tabular} & \begin{tabular}[c]{@{}l@{}}\#Words \\ \#Sentences\end{tabular} & \multicolumn{1}{c|}{\begin{tabular}[c]{@{}c@{}}0.04\\ 0.05\end{tabular}} & \multicolumn{1}{c|}{\begin{tabular}[c]{@{}c@{}}\textbf{-0.42**}\\ \textbf{-0.41*}\end{tabular}} & \begin{tabular}[c]{@{}c@{}}1\textless{}50\textless{}13893\\ 1\textless{}3\textless{}699\end{tabular} & \multicolumn{1}{c|}{\begin{tabular}[c]{@{}c@{}}0.99\\ 0.99\end{tabular}} & \multicolumn{1}{c|}{\begin{tabular}[c]{@{}c@{}}0.01\\ -0.01\end{tabular}} & \begin{tabular}[c]{@{}c@{}}1\textless{}397\textless{}13374\\ 1\textless{}23\textless{}809\end{tabular} \\ \hline
\end{tabular}%
}
\begin{tablenotes}
\footnotesize
\item\(^{*}\) Indicate differences that are statistically significant with (\(p < 0.05\)).
\item\(^{**}\) Indicate differences that are statistically significant with both (\(p < 0.05\)) and (\(corrected\ p < 0.05\))
\end{tablenotes}
\end{table*}

\subsubsection{Conversational Metrics} We present the results of \textit{Structural} and \textit{Linguistic} metrics in Table \ref{tab:rq2_1}. These metrics are calculated separately for the developer prompts and ChatGPT responses, as shown in columns (\textit{Developers}) and (\textit{ChatGPT}), respectively. For each row in this table, we show the relative difference (\textit{Rel. Diff.}) by calculating the mean average of the helpful (h1) and unhelpful (h2), and then determining the difference as 
$(h1 - h2) / h1$ between helpful and unhelpful conversations. 
We bold those that are statistically significant according to the Mann-Whitney U test (p-value\textless 0.05) \citep{Mann-Whitney}. The Mann-Whitney U test does not assume normal data distribution and is less affected by outliers and skewed data compared to parametric tests \citep{hollander2013nonparametric}, making it suitable for our study. We first discuss the results for the metrics in \textit{developer prompts}. We also present corrected p-values using the Benjamini-Hochberg (BH) correction~\citep{Benjamini} as an additional reliability measure, to show how many of the differences remain statistically significant after adjustment.
We use the BH correction to control the false discovery rate (FDR), reducing the risk of false positives among significant results. With many metrics analyzed, this helps ensure the validity of our findings despite the increased likelihood of false positives from multiple comparisons. In multiple tests, the probability of falsely rejecting at least one null hypothesis (Type I error) increases. This correction adjusts p-values to account for the number of tests, reducing this risk. Unlike stricter corrections like Bonferroni~\citep{armstrong_when_2014}, which control the family-wise error rate and can be overly conservative, the BH correction controls the FDR. This allows for more discoveries while maintaining a low proportion of false positives among significant results~\cite{Banerjee23}. Corrected p-values ensure our findings are less likely due to random chance.
Of the 17 metrics initially found to be significant, 11 remained statistically significant after adjustment, indicating that these metrics are robust indicators and less likely to be false positives. 

\noindent
\textbf{Knowledge Seeking/Sharing.} In developer prompts, the \textit{\#Primary Questions} is higher in helpful conversations (0.25), while the \textit{\#Knowledge-seeking Questions} is higher in unhelpful ones (-0.02). However, neither of these measures was statistically significant. We observed that 
unhelpful conversations contain more clarifying questions. 
For instance, in one conversation, a developer struggled to find instructions for downloading a model from a website, stating: \textit{``I'm trying to download this AI from hugging face and I can't find any explanation online anywhere..."}~\citep{knowledge_seeking}. 
After four rounds of back-and-forth interaction, looking for more clarification, the developer asked:  \textit{``do I absolutely NEED to do the virtual env? Can I just start with pip install transformers? normal answer"} and followed up with this knowledge-seeking question: \textit{``I don't have pytorch at all just python and pip. how can I install pytorch"}. The conversation continued with more questions about verifying the installation. 
This pattern explains why unhelpful conversations tend to have more knowledge-seeking questions.

\noindent
\textbf{Contextual. } 
\textit{\#Code Snippets} and \textit{\#Error Messages} in developer prompts are higher in helpful conversations, however, the \textit{\#Mean Size Code Snippets} is larger in unhelpful conversations, with a relative difference of -0.13. Providing very large code snippets is a common pattern observed in unhelpful conversations in our dataset. This suggests that ChatGPT struggles with handling large code snippets.
\textit{\#API Mentions} and \textit{\#Code Descriptions} are statistically higher in unhelpful conversations, where developers often refer to unclear or incorrect code suggestions made by ChatGPT.
The \textit{\#URLs} in developer prompts is higher in helpful conversations. Providing additional sources of information could be beneficial for issue resolution. For example, in a helpful conversation, a developer provided a link to help ChatGPT understand the context of the provided code: \textit{``Consider [URL]: [CODE] We would like to read frames in memory, not save to file"}~\citep{contextual}. However, in unhelpful conversations, we observed a few instances where ChatGPT was unfamiliar with the provided links.
This highlights a limitation of ChatGPT and other LLMs regarding data training, as their knowledge is capped at a certain date.

\noindent
\textbf{Psycholinguistics. }All measures are statistically significant and different in developer prompts between helpful and unhelpful conversations. The occurrence of words associated with politeness (\textit{\#Polite Words}), positive sentiment (\textit{\#Positive Tone Words}), and affection (\textit{\#Affect Words}) is notably higher in helpful conversations.
This suggests that a positive and polite tone in developer prompts is associated with successful issue resolution. 
In unhelpful conversations, instances of frustration with ChatGPT's responses were observed, with developers expressing their discontent through messages like \textit{``the zip file name is wrong dimwit"}~\citep{toxic}. 
Unhelpful conversations contain more words related to \textit{causality}, \textit{insight}, and \textit{cognitive} process. This suggests that when developers cannot resolve an issue with ChatGPT, they engage in more complex and analytical thinking. They use more causal, insight, and cognitive words to clarify their issues and make their thought process transparent, hoping for better responses from ChatGPT \citep{pennebaker_psychological_2003, Clark1991GroundingIC}.

\noindent
\textbf{Diversity. }Unhelpful conversations exhibit a significantly higher number of \textit{\#Unique Words} and \textit{\#Unique Info}, compared to helpful ones. 
LLMs have limitations on the amount of context and tokens they can process~\citep{10.1145/3649506}. Providing too much information can be overloading to the LLM, causing confusion about where to direct its attention. 
We noticed that developers asking for assistance with different tasks within the same conversation thread often led to unsuccessful outcomes.  

\noindent
\textbf{Readability. }In developer prompts, unhelpful conversations have more \textit{\#Misspellings} and \textit{\#Incomplete Sentences}. Additionally, both  \textit{Flesch Reading Ease Score} and \textit{SMOG Grade} are higher in prompts of helpful conversations. These scores indicate that developers' prompts in helpful conversations are well-written and easier to comprehend.


\noindent
\textbf{Verbosity. }The \textit{\#Words} and \textit{\#Sentences} are both significantly higher in prompts sent in unhelpful conversations. These findings align with prior studies \citep{hao2024empirical}, which indicate that developers often engage in iterative back-and-forth loops with ChatGPT, seeking clarification, follow-ups, and prompt refinements to find answers. These unsuccessful loops are prevalent in unhelpful conversations, leading to the increased number of words and sentences within these conversations. \textcolor{black}{As shown in Table~\ref{tab:dataset2}, the distribution of prompt counts across helpful and unhelpful conversations is nearly identical, with 93\% of conversations in both groups involving 0–10 prompts. This suggests that the number of prompts alone does not distinguish helpfulness; rather, it is the efficiency of the exchanges that matters. Helpful conversations may include multiple rounds, but they tend to be more concise, enabling developers to iterate toward solutions with fewer words and sentences overall.}

Structural and Linguistic metrics in the \textit{ChatGPT responses} indicate only one feature with a statistically significant difference between helpful and unhelpful conversations: \textit{\#Code Snippets}, with a relative difference of 0.28.
ChatGPT's answers in helpful conversations contained more code snippets, driving effective issue resolution. The other features showed minimal differences, suggesting that ChatGPT maintains a consistent tone and language style across its conversations with developers, in both helpful and unhelpful conversations.

\begin{table}[h]
\caption{\textcolor{black}{Sentiment and Prompt Statistics for Helpful vs. Unhelpful Conversations}}
\label{tab:dataset2}
\resizebox{\columnwidth}{!}{%
\begin{tabular}{c|ccc|c|c|c|cccc}
\multirow{2}{*}{\textbf{\begin{tabular}[c]{@{}c@{}}ChatGPT\\ Conversations\end{tabular}}} & \multicolumn{3}{c|}{\textbf{Sentiment}} & \multirow{2}{*}{\textbf{\#Words}} & \multirow{2}{*}{\textbf{\#Sentences}} & \multirow{2}{*}{\textbf{\begin{tabular}[c]{@{}c@{}}Median \\ \#Prompts\end{tabular}}} & \multicolumn{4}{c}{\textbf{Rounds of Conversation}} \\ \cline{2-4} \cline{8-11} 
 & \multicolumn{1}{c|}{\textbf{Neg.}} & \multicolumn{1}{c|}{\textbf{Neut.}} & \textbf{Pos.} &  &  &  & \multicolumn{1}{c|}{\textbf{0 to 10}} & \multicolumn{1}{c|}{\textbf{10 to 20}} & \multicolumn{1}{c|}{\textbf{20 to 25}} & \textbf{25\textgreater{}} \\ \hline
\textbf{\begin{tabular}[c]{@{}c@{}}Helpful\\ (n=428)\end{tabular}} & \multicolumn{1}{c|}{\begin{tabular}[c]{@{}c@{}}70\\ (16\%)\end{tabular}} & \multicolumn{1}{c|}{\begin{tabular}[c]{@{}c@{}}284\\ (66\%)\end{tabular}} & \begin{tabular}[c]{@{}c@{}}75\\ (18\%)\end{tabular} & 61348 & 1861 & 2 & \multicolumn{1}{c|}{\begin{tabular}[c]{@{}c@{}}401\\ (93\%)\end{tabular}} & \multicolumn{1}{c|}{\begin{tabular}[c]{@{}c@{}}20\\ (4\%)\end{tabular}} & \multicolumn{1}{c|}{\begin{tabular}[c]{@{}c@{}}4\\ ($\sim$1\%)\end{tabular}} & \begin{tabular}[c]{@{}c@{}}3\\ ($\sim$1\%)\end{tabular} \\ \hline
\textbf{\begin{tabular}[c]{@{}c@{}}Unhelpful\\ (n=258)\end{tabular}} & \multicolumn{1}{c|}{\begin{tabular}[c]{@{}c@{}}38\\ (15\%)\end{tabular}} & \multicolumn{1}{c|}{\begin{tabular}[c]{@{}c@{}}192\\ (74\%)\end{tabular}} & \begin{tabular}[c]{@{}c@{}}28\\ (11\%)\end{tabular} & 106476 & 3558 & 2 & \multicolumn{1}{c|}{\begin{tabular}[c]{@{}c@{}}241\\ (93\%)\end{tabular}} & \multicolumn{1}{c|}{\begin{tabular}[c]{@{}c@{}}11\\ (4\%)\end{tabular}} & \multicolumn{1}{c|}{\begin{tabular}[c]{@{}c@{}}5\\ ($\sim$1\%)\end{tabular}} & \begin{tabular}[c]{@{}c@{}}1\\ ($\sim$1\%)\end{tabular}
\end{tabular}%
}
\vspace{-0.1cm}
\end{table}

\begin{figure}[ht!]
    \centering
    \includegraphics[width=0.45\linewidth]{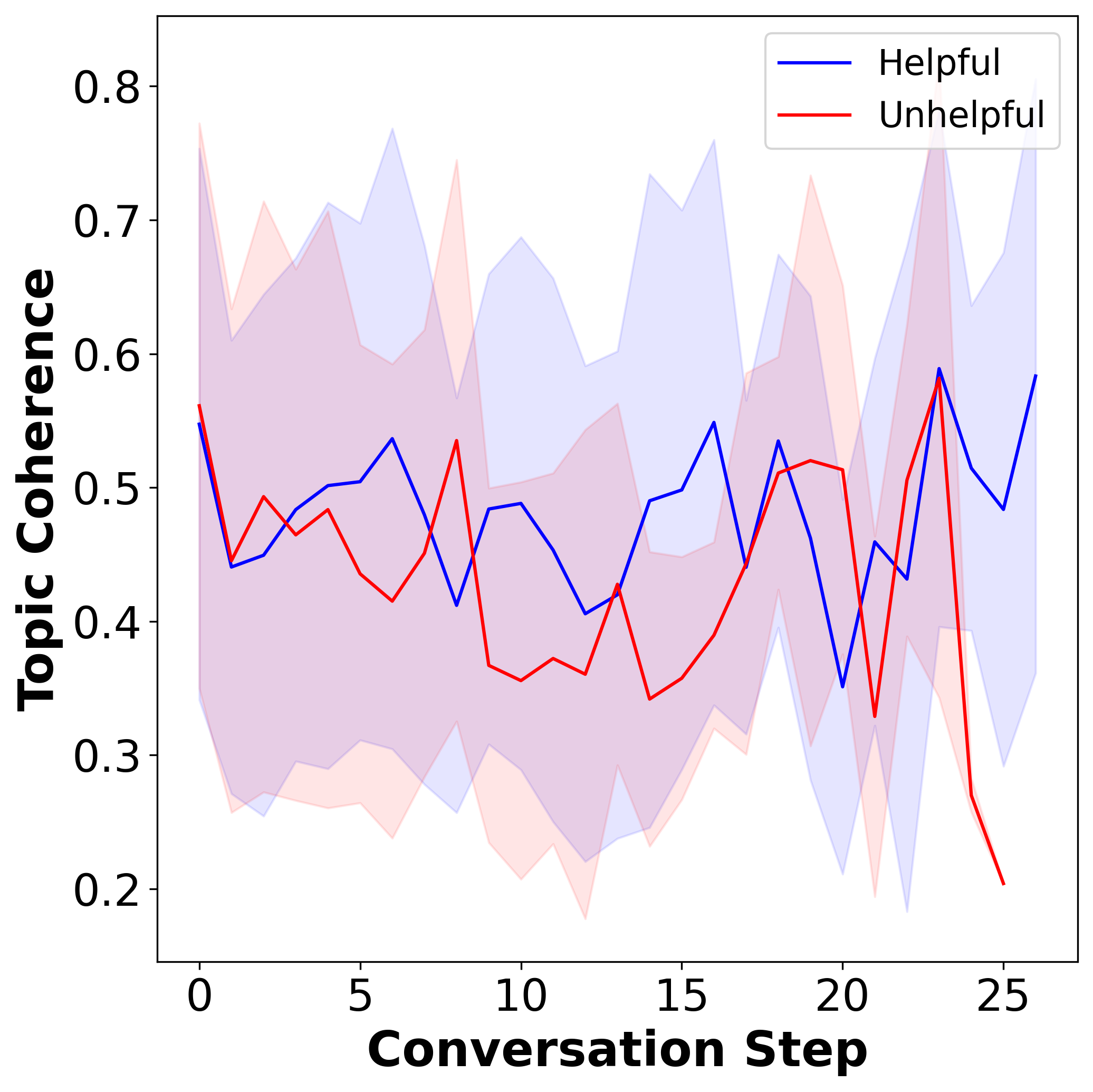}
    \caption{Topic Coherence Measured for Helpful/Unhelpful Conversations.}
    \label{fig:coherence}
\end{figure}

\begin{figure}[ht!]
    \centering
    \includegraphics[width=0.45\linewidth]{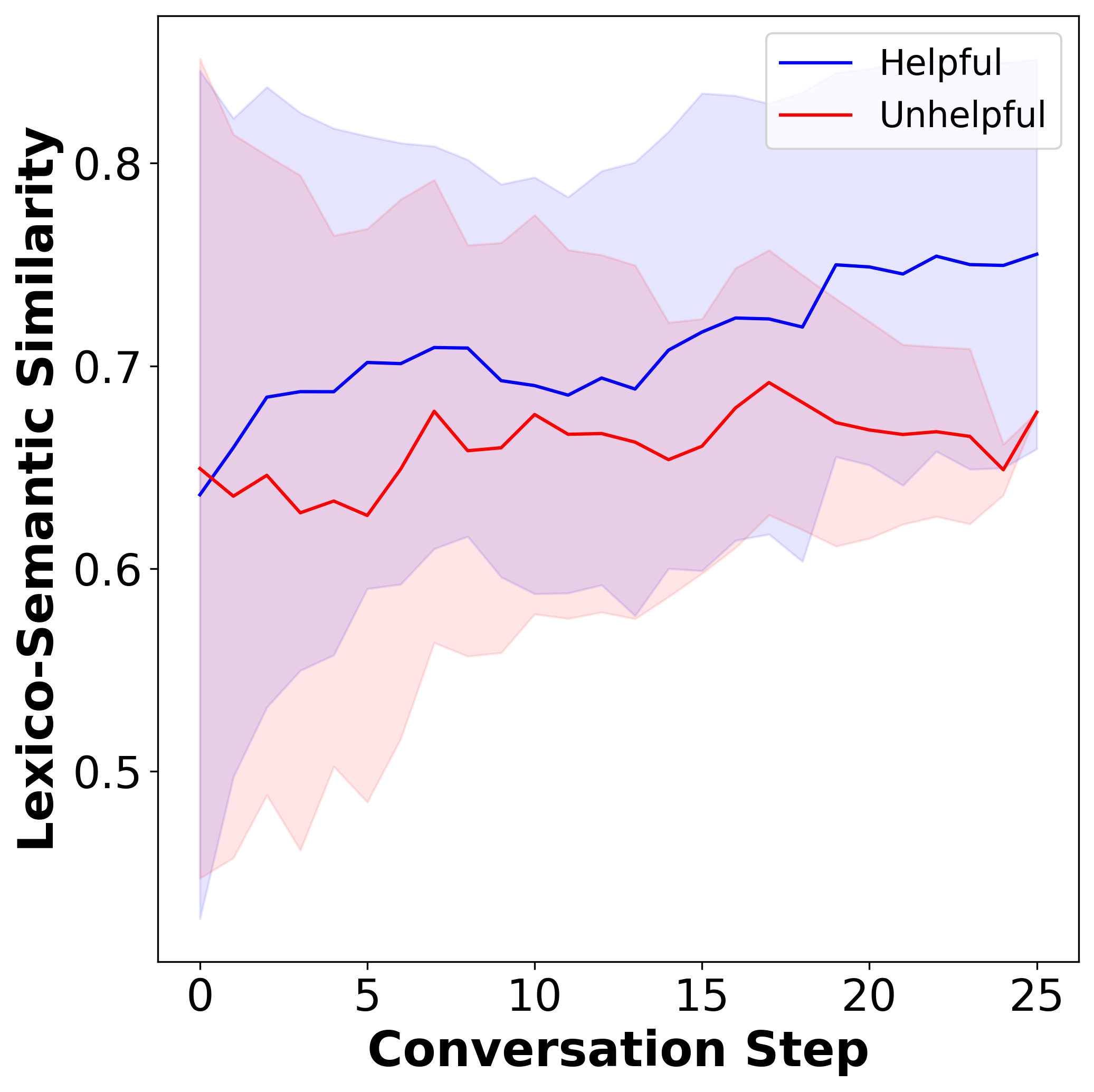}
    \caption{Lexico-Semantic Similarity Measured for Helpful/Unhelpful Conversations.}
    \label{fig:cosine_similarity}
\end{figure}

\begin{figure}[ht!]
    \centering
    \includegraphics[width=0.45\linewidth]{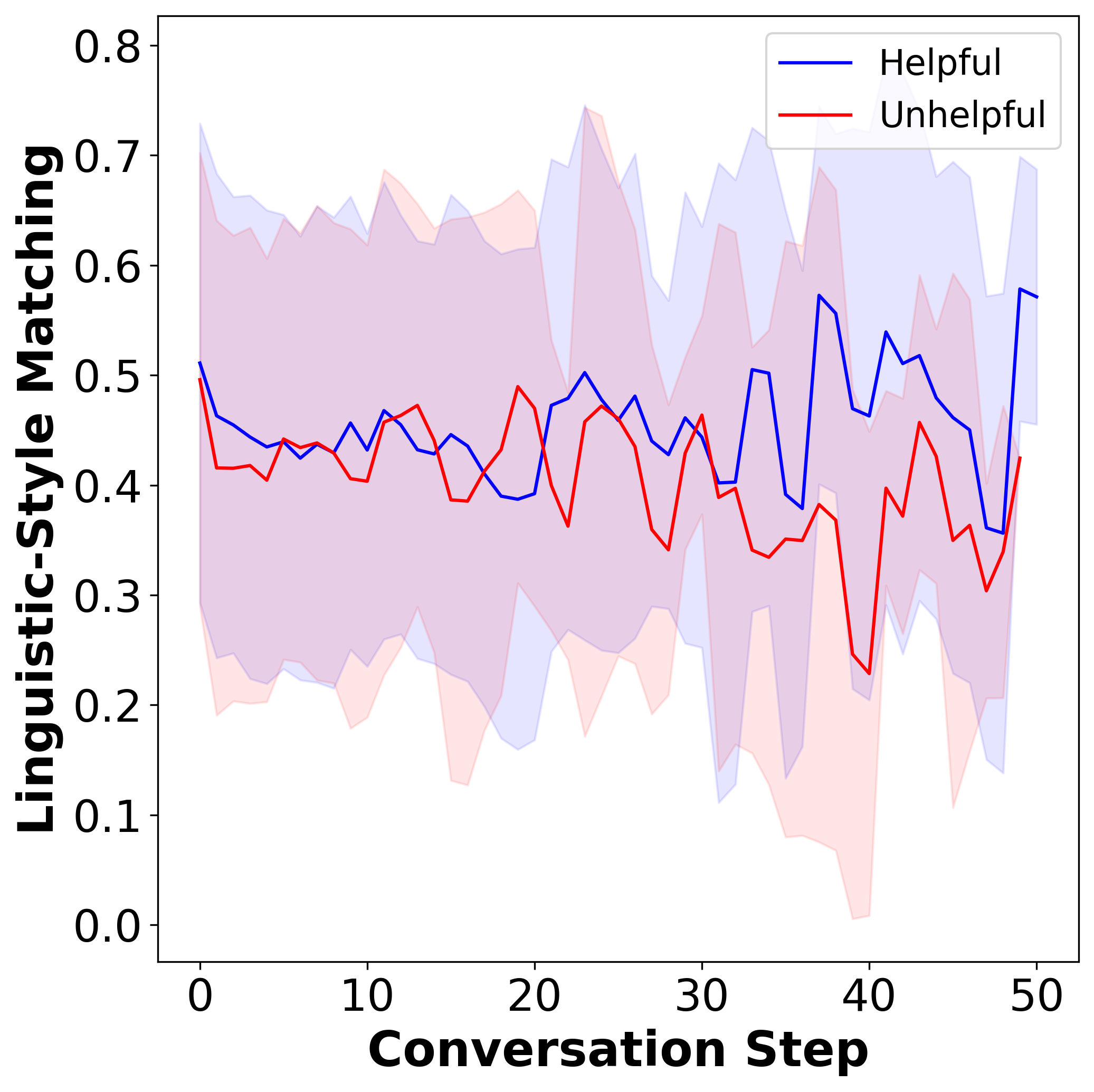}
    \caption{Linguistic-Style Matching Measured for Helpful/Unhelpful Conversations.}
    \label{fig:lsm}
\end{figure}

We present the results of \textit{Discourse} metrics in Figures \ref{fig:coherence}, \ref{fig:cosine_similarity}, and \ref{fig:lsm}. 
In each figure, we plot the average scores for each metric (Topic Coherence, Lexico Semantic Similarity, Linguistic Style Matching). 
In these figures, A ``conversation step" refers to each round of interaction, consisting of one developer prompt and one ChatGPT response, and the shaded areas in the plots represent the standard deviation of the metrics in helpful (blue) and unhelpful (red) conversations throughout the discourse.
\textcolor{black}{In Table~\ref{tab:dataset2}, we provide descriptive statistics on conversation rounds, word and sentence counts, and user sentiment for both helpful and unhelpful conversations. The distributions of conversation rounds are nearly identical across the two groups. For instance, the majority of the conversations occur within 0 to 10 rounds (401 out of 428 helpful conversations, 241 out of 258 conversations)}. 

\textcolor{black}{The sentiment patterns also follow a similar trend, with the majority of prompts being neutral in both helpful and unhelpful conversations. However, helpful conversations exhibit a greater proportion of positive sentiment (18\%) compared to unhelpful ones (11\%), indicating that more positively framed conversations tend to be perceived as more helpful. Prompting with phrases such as ``\textit{I'd love for you to point me to some examples}" or expressions of gratitude are examples of such positive conversations~\citep{issue_type_9}.}

\noindent
\textbf{Topic Coherence. }Topic coherence scores are generally higher in helpful conversations compared to unhelpful ones, as shown in Figure \ref{fig:coherence}. 
Drops in scores signify topic deviations between consecutive steps (e.g., noticeable drop between steps 8 and 9 in unhelpful conversations). One such instance was where a developer first sought help with a specific error: \textit{``(PY0001): PyScript: Access to local files...[ERROR]. See [URL]..."}~\citep{shift_conversation}, to which ChatGPT provided potential solutions. Immediately at the next step, the developer abruptly shifted to a different topic: \textit{``how to put python code in website"}. These abrupt topic changes were more common in unhelpful conversations. 

\noindent
\textbf{Lexico-Semantic Similarity. }As shown in Figure \ref{fig:cosine_similarity}, lexico-semantic similarity increases for both helpful and unhelpful conversations, indicating that developer prompts and ChatGPT responses become more semantically aligned as the conversation progresses. However, helpful conversations consistently show higher similarity. Our analysis reveals that this score decreases when developers include error messages and code snippets without accompanying explanations or specific requests, a pattern frequently observed in unhelpful conversations. Score drops also occur due to misalignment in understanding the developer's request. For example, in an unhelpful conversation~\citep{Lexico}, the developer initially asked: \textit{``create load generator in go"}, to which ChatGPT provided an unrelated answer that did not meet the developer's needs. The developer then clarified with: \textit{``I mean CPU load"}, highlighting the misalignment. These kinds of misunderstandings also contribute to score drops, which are more common in unhelpful conversations.

\noindent
\textbf{Linguistic-Style Matching (LSM). }The linguistic styles of developers and ChatGPT are generally similar in both helpful and unhelpful conversations, as shown in Figure \ref{fig:lsm}. However, this trend shifts at step 32 with helpful conversations showing greater linguistic alignment. Helpful conversations generally involve developers making an effort to structure their prompts clearly, facilitating ChatGPT's understanding and providing better assistance. For instance, in a high-scoring LSM conversation, a developer tries to adjust an application feature with ChatGPT's help. Throughout the interaction, the developer tries the scripts provided by ChatGPT and gives clear feedback on the results~\citep{LSM}: \textit{``It seems there was an error in the script, and it appears that the [CODE] library does not have a constant named [CODE]"}, or \textit{``this is very close. I am wondering why it is moving the left keyboard arrow"}. This clear communication ultimately leads to resolving the issues with the script.

\subsubsection{Project Metrics} We present the results in Table \ref{tab:rq2_2}.
For each row in this table, we show the relative difference (Rel. Diff.) between the projects involved in helpful and unhelpful conversations, bolding those that are statistically significant according to the Mann-Whitney U test (p-value\textless 0.05) \citep{Mann-Whitney}. We also add Benjamini-Hochberg corrected values in the table. 

\noindent
\textbf{Repository Parameters. }
The \textit{\#Files} and \textit{\#Lines of Code} are both statistically significant and higher in helpful conversations, indicating that projects with extensive codebases tend to have more effective interactions with ChatGPT. Additionally, \textit{\#Stars} is also higher in helpful conversations, suggesting that popular projects benefit more from ChatGPT's assistance. Conversely, projects associated with unhelpful conversations have slightly more contributors and forks, although they are not statistically significant. 

\begin{table}[]
\caption{Relative Differences Between Helpful and Unhelpful Developer-ChatGPT Conversations for Project Metrics. {\textbf{Positive}} Numbers Indicate Higher Frequencies of Metrics, and {\textbf{Negative}} Numbers Indicate Lower Frequencies in Helpful Conversations.}
\label{tab:rq2_2}
\resizebox{\columnwidth}{!}{%
\begin{tabular}{|c|l|c|c|c|c|}
\hline
\textbf{\begin{tabular}[c]{@{}c@{}}Project \\ Metric\end{tabular}} & \textbf{Categories} & \textbf{P-value} & \textbf{\begin{tabular}[c]{@{}c@{}}Corrected\\ P-Value\end{tabular}} & \textbf{Rel. Diff.} & \textbf{\begin{tabular}[c]{@{}c@{}}Range\\ (Min\textless Median\textless Max)\end{tabular}} \\ \hline
\begin{tabular}[c]{@{}c@{}}Repository\\ Parameters\end{tabular} & \begin{tabular}[c]{@{}l@{}}\#Stars\\ \#Contributors\\ \#Forks\\ \#Files\\ \#Lines of Code\end{tabular} & \begin{tabular}[c]{@{}c@{}}0.73\\ 0.39\\ 0.83\\ 0.0002\\ 0.0008\end{tabular} & \begin{tabular}[c]{@{}c@{}}0.91\\ 0.65\\ 0.83\\ 0.001\\ 0.002\end{tabular} & \begin{tabular}[c]{@{}c@{}}{0.27}\\ {-0.01}\\ {-0.01}\\ {\textbf{0.40**}}\\ {\textbf{0.72**}}\end{tabular} & \begin{tabular}[c]{@{}c@{}}0\textless23\textless116,455\\ 1\textless5\textless30\\ 0\textless6\textless35,443\\ 0\textless85\textless48,122\\ 0\textless12,936\textless103,840,135\end{tabular} \\ \hline
\begin{tabular}[c]{@{}c@{}}Developer\\ Experience\end{tabular} & \begin{tabular}[c]{@{}l@{}}\#Public Repositories\\ \#Followers\\ Account Age\\ \#Contributions\end{tabular} & \begin{tabular}[c]{@{}c@{}}0.18\\ 0.10\\ 0.05\\ 0.03\end{tabular} & \begin{tabular}[c]{@{}c@{}}0.18\\ 0.13\\ 0.10\\ 0.12\end{tabular} & \begin{tabular}[c]{@{}c@{}}{0.18}\\ {0.24}\\ {0.10}\\ {\textbf{0.05*}}\end{tabular} & \begin{tabular}[c]{@{}c@{}}0\textless42\textless8,728\\ 0\textless27\textless42,042\\ 0\textless9\textless16\\ 0\textless20\textless30\end{tabular} \\ \hline
\end{tabular}%
}
\begin{tablenotes}
\footnotesize
\item\(^{*}\) Indicate differences that are statistically significant with (\(p < 0.05\)).
\item\(^{**}\) Indicate differences that are statistically significant with both (\(p < 0.05\)) and (\(corrected\ p < 0.05\))
\end{tablenotes}
\end{table}

\noindent
\textbf{Developer Experience. }All measures in this category are higher in helpful conversations, with \textit{\#Contributions} being statistically significant. \textcolor{black}{This suggests that effectively leveraging ChatGPT for issue resolution is not solely about access to the tool; it also depends on the developer’s experience and familiarity with software development workflows. Conversations between experienced developers and ChatGPT are more likely to be helpful compared to those involving less experienced developers. This is especially important in the context of the new trend of \textit{vibe coding}, where less experienced developers may rely heavily on LLMs without fully understanding the system. Our findings suggest that such reliance, without sufficient experience, increases the risk of unhelpful or misleading outcomes.
Experienced developers tend to provide more context, clearer questions, and more relevant information, which likely leads to better responses from ChatGPT. Additionally, experienced developers are better at interpreting ChatGPT's responses and providing follow-up queries or corrections, further enhancing the conversation's productivity and helpfulness. This highlights the critical role of experience in effectively using ChatGPT for problem-solving and issue resolution.}

\subsubsection{\textcolor{black}{Issue Metrics}}
\textcolor{black}{We present the results for Issue Difficulty in Table~\ref{tab:rq2_3}. For each row, the table reports the relative difference (Rel. Diff.) between issue threads associated with helpful and unhelpful conversations. Rows that are statistically significant based on the Mann–Whitney U test (p-value\textless0.05) are shown in bold. We also include Benjamini–Hochberg corrected p-values to account for multiple comparisons and reduce the likelihood of false positives.}

\begin{table}[h!]
\caption{\textcolor{black}{Relative Differences Between Helpful and Unhelpful Developer-ChatGPT Conversations for Issue Difficulty. {\textbf{Positive}} Numbers Indicate Higher Frequencies of Difficult Issues, and {\textbf{Negative}} Numbers Indicate Lower Frequencies in Helpful Conversations.}}
\label{tab:rq2_3}
\resizebox{\columnwidth}{!}{%
\begin{tabular}{|c|l|c|c|c|c|}
\hline
\textbf{\begin{tabular}[c]{@{}c@{}}Issue \\ Metric\end{tabular}} & \textbf{Categories} & \textbf{P-value} & \textbf{\begin{tabular}[c]{@{}c@{}}Corrected\\ P-Value\end{tabular}} & \textbf{Rel. Diff.} & \textbf{\begin{tabular}[c]{@{}c@{}}Range\\ (Min\textless{}Median\textless{}Max)\end{tabular}} \\ \hline
Issue Difficulty & \begin{tabular}[c]{@{}l@{}}\#Comments\\ Length of Discussion\\ \#Discussion Hours\\ \#Unique Developers\\ Resolution time\\ Patch Size\end{tabular} & \begin{tabular}[c]{@{}c@{}}0.82\\ 0.54\\ 0.93\\ 0.001\\ 0.001\\ 0.001\end{tabular} & \begin{tabular}[c]{@{}c@{}}0.96\\ 0.76\\ 0.93\\ 0.002\\ 0.002\\ 0.002\end{tabular} & \begin{tabular}[c]{@{}c@{}}-0.16\\ -0.42\\ 0.002\\ \textbf{-0.11**}\\ \textbf{-0.002**}\\ \textbf{0.70**}\end{tabular} & \begin{tabular}[c]{@{}c@{}}0\textless{}3\textless{}30\\ 0\textless{}77\textless{}8378\\ 0\textless{}238\textless{}68591\\ 1\textless{}2\textless{}24\\ 0\textless{}294\textless{}68591\\ 0\textless{}0\textless{}26718\end{tabular} \\ \hline
\end{tabular}%
}
\begin{tablenotes}
\footnotesize
\item\(^{*}\) Indicate differences that are statistically significant with (\(p < 0.05\)).
\item\(^{**}\) Indicate differences that are statistically significant with both (\(p < 0.05\)) and (\(corrected\ p < 0.05\))
\end{tablenotes}
\end{table}

\textcolor{black}{
Table~\ref{tab:rq2_4} presents the distribution of issue types identified for each conversation.
We identified five distinct issue types across the dataset:
\begin{itemize}
    \item \textit{API/Library Feature Requests}, such as issues requesting new functionality, methods, or improvements to existing APIs and libraries.
    \item \textit{Installation and Configuration Issues}, problems related to project setup, environment configuration, dependency management, and onboarding.
    \item \textit{Compatibility and Compilation Errors}, characterized by errors stemming from version mismatches, platform incompatibility, or build failures.
    \item \textit{Performance Optimization and Code Refactoring}, involving reports focused on slow runtime, high memory usage, or suggestions for improving code efficiency and structure.
    \item \textit{Debugging and Testing Issues}, including bugs encountered during development, test failures, and issues involving mocking, test coverage, or flaky tests.
\end{itemize}
The table reports the frequency of each issue type and its distribution across helpful and unhelpful conversations.}

\begin{table}[h!]
\caption{\textcolor{black}{Identified Issue Types for Helpful and Unhelpful Conversations and their Frequencies}}
\label{tab:rq2_4}
\resizebox{\columnwidth}{!}{%
\begin{tabular}{|l|c|c|c|}
\hline
\multicolumn{1}{|c|}{\textbf{Issue Type}} & \textbf{\begin{tabular}[c]{@{}c@{}}Total\\ Conversations\end{tabular}} & \textbf{\begin{tabular}[c]{@{}c@{}}Helpful\\ Conversations\end{tabular}} & \textbf{\begin{tabular}[c]{@{}c@{}}Unhelpful\\ Conversations\end{tabular}} \\ \hline
\begin{tabular}[c]{@{}l@{}}Debugging and\\ Testing Issues\end{tabular} & 223 & 134 (60\%) & \textbf{89 (40\%)} \\ \hline
\begin{tabular}[c]{@{}l@{}}API and Library\\ Feature Requests\end{tabular} & 192 & \textbf{125 (65\%)} & 67 (35\%) \\ \hline
\begin{tabular}[c]{@{}l@{}}Installation and\\ Configuration Issues\end{tabular} & 98 & \textbf{63 (64\%)} & 35 (36\%) \\ \hline
\begin{tabular}[c]{@{}l@{}}Performance Optimization\\ and Code Refactoring\end{tabular} & 87 & 48 (55\%) & \textbf{39 (45\%)} \\ \hline
\begin{tabular}[c]{@{}l@{}}Compatibility and\\ Compilation Errors\end{tabular} & 86 & \textbf{58 (67\%)} & 28 (33\%) \\ \hline
\end{tabular}%
}
\end{table}

\textcolor{black}{These issue types also align closely with the tasks in the conversations we identified in RQ1. For example, \textit{Debugging and Testing Issues} often involve tasks like Bug Identification and Fixing, while \textit{API/Library Feature Requests} correspond to Tool/ Library/API Recommendation. Similarly, \textit{Compatibility and Compilation Errors} and \textit{Installation and Configuration Issues} often prompt developers to seek Code Generation or implementation help. This overlap reinforces that the issue types reflect the nature of the developer–ChatGPT conversations in our dataset.}

\noindent
\textcolor{black}{\textbf{Issue Difficulty.} Unhelpful conversations are more frequently associated with issue threads that show higher engagement from the developers, reflected in a greater number of comments (\textit{\#Comments}), longer discussions (\textit{Length of Discussion}), and more participating developers (\textit{\#Unique Developers}). Among these, only \textit{\#Unique Developers} is statistically significant. Additionally, unhelpful conversations are associated with significantly longer resolution times (\textit{Resolution Time}). These findings suggest that unhelpful conversations are more likely to occur in complex, collaborative, and time-consuming issues. Coordination across multiple contributors reflects more complex or ambiguous issues where ChatGPT struggles to provide actionable help. On the other hand, the only metric significantly higher in helpful conversations is \textit{Patch Size}, indicating that when ChatGPT is helpful, it contributes to larger and more substantial code changes by helping developers more confidently make broader updates or refactorings.}

\noindent
\textcolor{black}{\textbf{Issue Type. } ChatGPT is most helpful in resolving \textit{Compatibility and Compilation Errors} (67\% helpful), \textit{API and Library Feature Requests} (65\% helpful), and \textit{Installation and Configuration Issues} (64\% helpful). These types of issues often involve focused, well-scoped technical questions related to known APIs, dependencies, or environment setup, contexts where ChatGPT can provide actionable suggestions based on its pretraining. In contrast, ChatGPT is relatively less helpful for \textit{Performance Optimization and Code Refactoring} (45\% unhelpful) and especially for complex \textit{Debugging and Testing Issues} (40\% unhelpful), which may involve subtle logic errors or incomplete bug descriptions. These tasks often require a deeper understanding of the project’s runtime behavior, internal state, or debugging context, information that is often missing or underspecified in prompts. These findings reinforce a key limitation of general-purpose LLMs: their effectiveness decreases when issue resolution depends on nuanced, dynamic, or implicit project-specific information that is not directly provided in the conversation.}

\begin{tcolorbox}[]
\textit{\textbf{Conversational Metrics:}} 
Developer prompts in helpful conversations are more readable (SMOG Grade), polite (\#Polite Words), and coherent (Topic Coherence). In contrast, unhelpful conversations are more verbose (\#Words), overloaded with too much information (\#Unique Info) and contain longer code snippets (Mean Size Code Snippets).

\textit{\textbf{Project Metrics:}} Larger projects (\#Files) use ChatGPT more effectively for issue resolution, and experienced developers (\#Contributions) leverage ChatGPT's capabilities more efficiently.

\textcolor{black}{\textit{\textbf{Issue Metrics:}} ChatGPT is less helpful for complex, time-consuming issues (\textit{Resolution Time}) that involve high developer collaboration (\textit{\#Comments}, \textit{Length of Discussion}, \textit{\#Unique Developers}). It performs best on well-defined technical issues like compilation errors and feature requests, but struggles with context-heavy debugging and refactoring issues that require deeper, project-specific understanding.}

\end{tcolorbox}

\subsection{RQ3. What are the common deficiencies in ChatGPT's responses in unhelpful issue-resolution conversations? How do they vary across different tasks?}

We analyze 258 unhelpful developer-ChatGPT conversations to identify deficiencies in ChatGPT's responses. We categorize these deficiencies into five categories: incorrect information (\textit{Correctness}), insufficient or incomplete information (\textit{Comprehensiveness}), unclear information leading to confusion (\textit{Clarity}), outdated or irrelevant information (\textit{Relevance}), and fabricated or nonsensical responses (\textit{Hallucination}).

ChatGPT's answers were identified as \textit{Incorrect} in 54 cases, \textit{Not Comprehensive} in 39 cases, \textit{Unclear} in 14 cases, \textit{Irrelevant} in 19 cases, and \textit{Hallucination} in 11 cases. The remaining 121 conversations are unresolved issue threads without clear indications of any specific deficiencies. The distribution of deficiencies across different tasks is shown in Table \ref{tab:heatmap_rq3}.

\begin{table}[h]
\caption{Number of Deficiencies in Unhelpful Conversations by Task.}
\label{tab:heatmap_rq3}
\renewcommand{\arraystretch}{1.5} 
\resizebox{\columnwidth}{!}{%
\begin{tabular}{m{3.5cm}|c|c|c|c|c}
 & \textbf{Incorrect} & \textbf{Not Comprehensive} & \textbf{Unclear} & \textbf{Irrelevant} & \textbf{Hallucination} \\ \hline
\textbf{Code Generation and Implementation} & 12 & 7 & 0 & \textbf{6} & 1 \\ \hline
\textbf{Tool/Library/API Recommendation} & \textbf{23} & \textbf{8} & \textbf{5} & \textbf{8} & \textbf{5} \\ \hline
\textbf{Bug Identification and Fixing} & \textbf{13} & 3 & 1 & 3 & 1 \\ \hline
\textbf{SE Information-Seeking} & 0 & \textbf{15} & \textbf{5} & 4 & \textbf{4} \\ \hline
\textbf{Code Enhancement} & 5 & 3 & 2 & 2 & 0 \\ \hline
\textbf{System Design and Architecture} & 0 & 1 & 0 & 0 & 0 \\ \hline
\textbf{Code Explanation} & 3 & 5 & 1 & 1 & 0 \\ \hline
\textbf{Test Generation} & 0 & 0 & 0 & 0 & 0 \\ \hline
\textbf{Code Documentation/Comment Generation} & 0 & 0 & 0 & 0 & 0
\end{tabular}%
}
\end{table}

\begin{figure}[h!]
    \centering
    \includegraphics[width=0.65\textwidth]{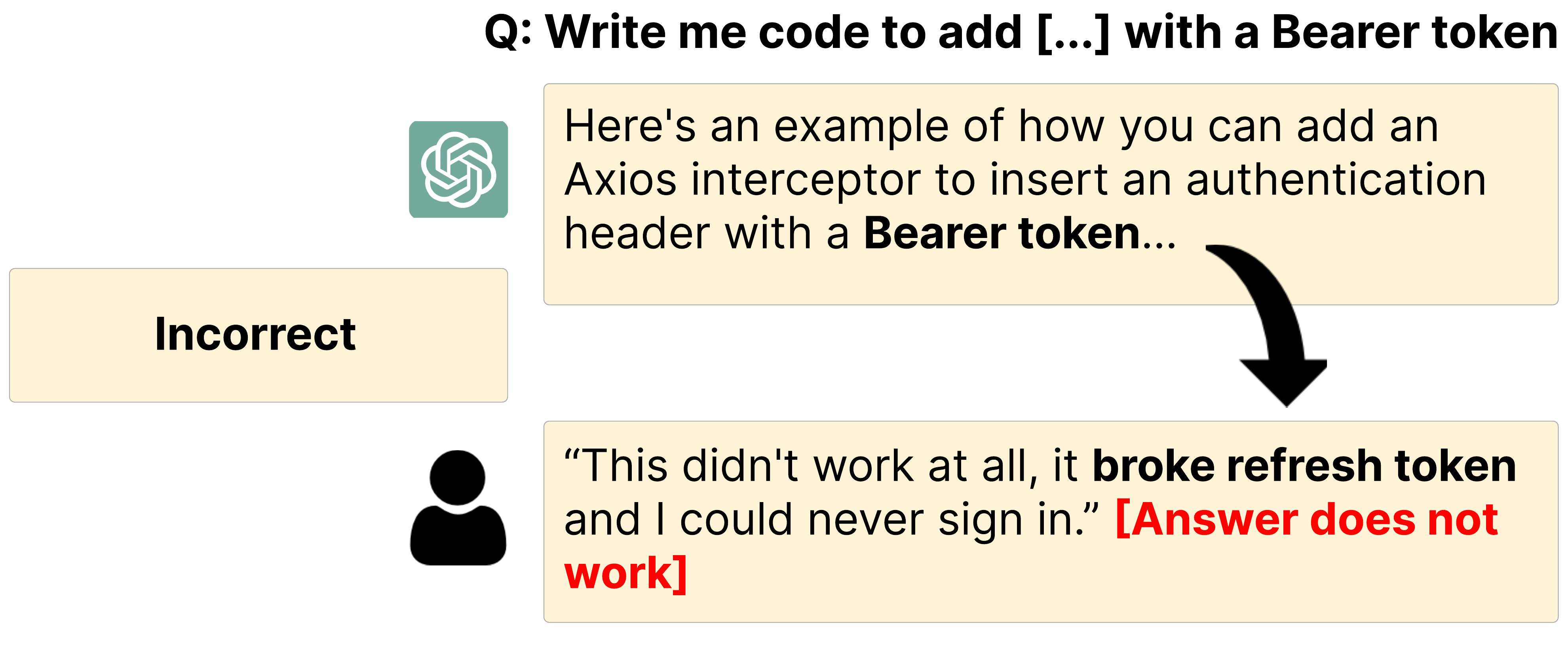}
    \caption{Example of Incorrect Deficiencies in ChatGPT's Answers.}
    \label{fig:examples_incorrect}
\end{figure}

\textit{Incorrect} answers are factually wrong or inaccurate responses, as reported by developers in the issue threads after testing or examining the provided solutions. For example, a developer mentioned, \textit{``I tried again [URL] and the regex it gave me is failing..."}~\citep{REGEX}. We also identified several instances where developers flagged the code provided by ChatGPT as bad practice, leading them to 
avoid using it for issue resolution (Figure \ref{fig:examples_incorrect}). The highest number of incorrect answers were noticed in the three most frequent tasks of \textit{Tool/Library/API Recommendation} (n=23), \textit{Bug Identification and Fixing} (n=13), and \textit{Code Generation and Implementation} (n=12). 

\begin{figure}[h]
    \centering
    \includegraphics[width=0.65\textwidth]{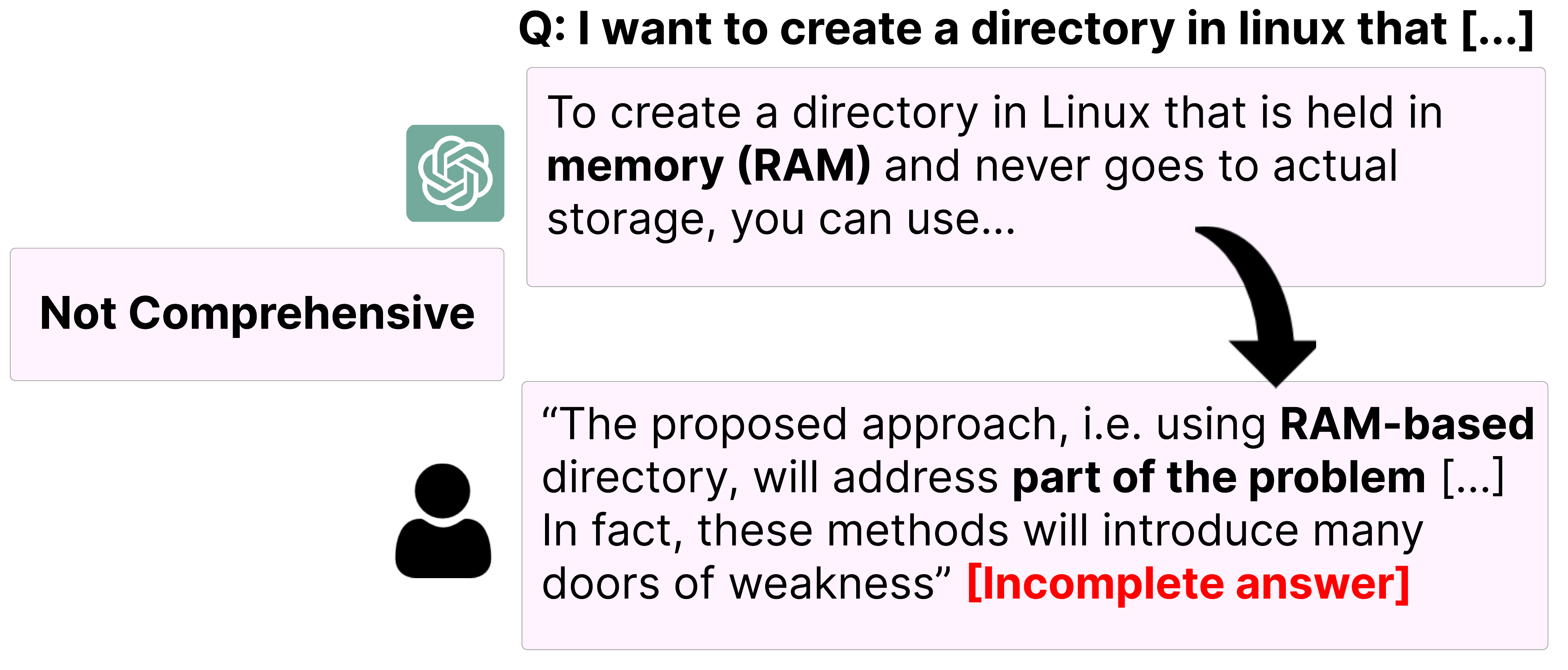}
    \caption{Example of Not Comprehensive Deficiencies in ChatGPT's Answers.}
    \label{fig:examples_comp}
\end{figure}

\textit{Not comprehensive} answers are responses that were found to be insufficient or incomplete for the developers.
For instance, one developer mentioned, \textit{``Hm, not too bad, but also still very generic..."}~\citep{generic}. Another example of this is shown in Figure \ref{fig:examples_comp}. Most instances of not comprehensive answers were observed in the tasks of \textit{SE Information-Seeking} (n=15), \textit{Tool/Library/API Recommendation} (n=8), and \textit{Code Generation and Implementation} (n=7), where the answers and information given by ChatGPT were too generic and incomplete to help resolve the issue.

\begin{figure}[h]
    \centering
    \includegraphics[width=0.65\textwidth]{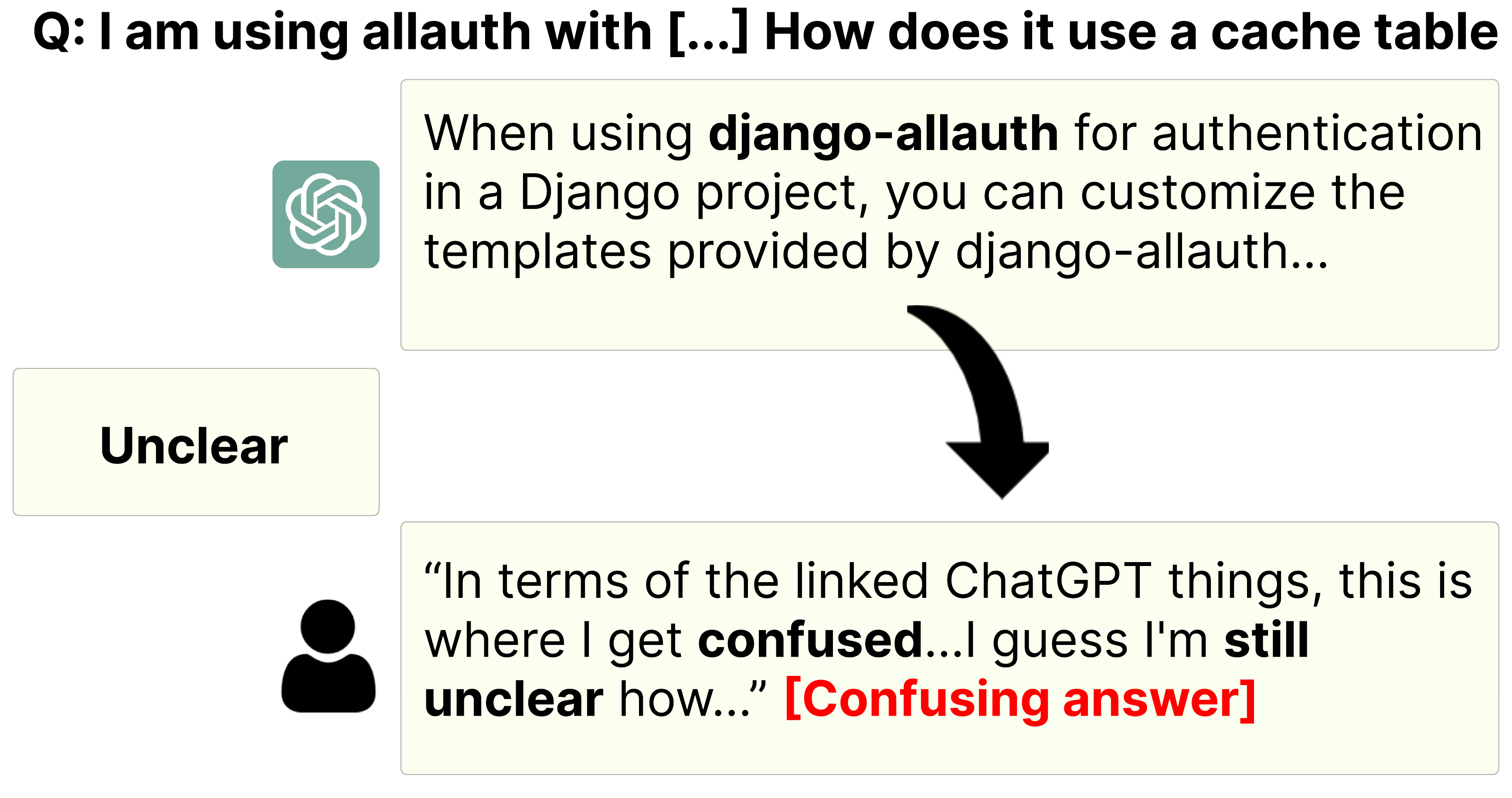}
    \caption{Example of Unclear Deficiencies in ChatGPT's Answers.}
    \label{fig:examples_clear}
\end{figure}

\textit{Unclear} answers are responses where the information by ChatGPT is vague and causes confusion among the developers (Figure \ref{fig:examples_clear}). The most frequent tasks with this deficiency are \textit{SE Information-Seeking} (n=5), \textit{Tool/Library/API Recommendation} (n=5), and \textit{Code Enhancement} (n=2). Across different tasks, we observed a recurring issue with the vagueness of ChatGPT's answers. For example, during a feature enhancement discussion, a developer noted, \textit{``[ChatGPT] got me absolute vagueries about the specifics"}~\citep{vague_answer}.

\begin{figure}[h]
    \centering
    \includegraphics[width=0.65\textwidth]{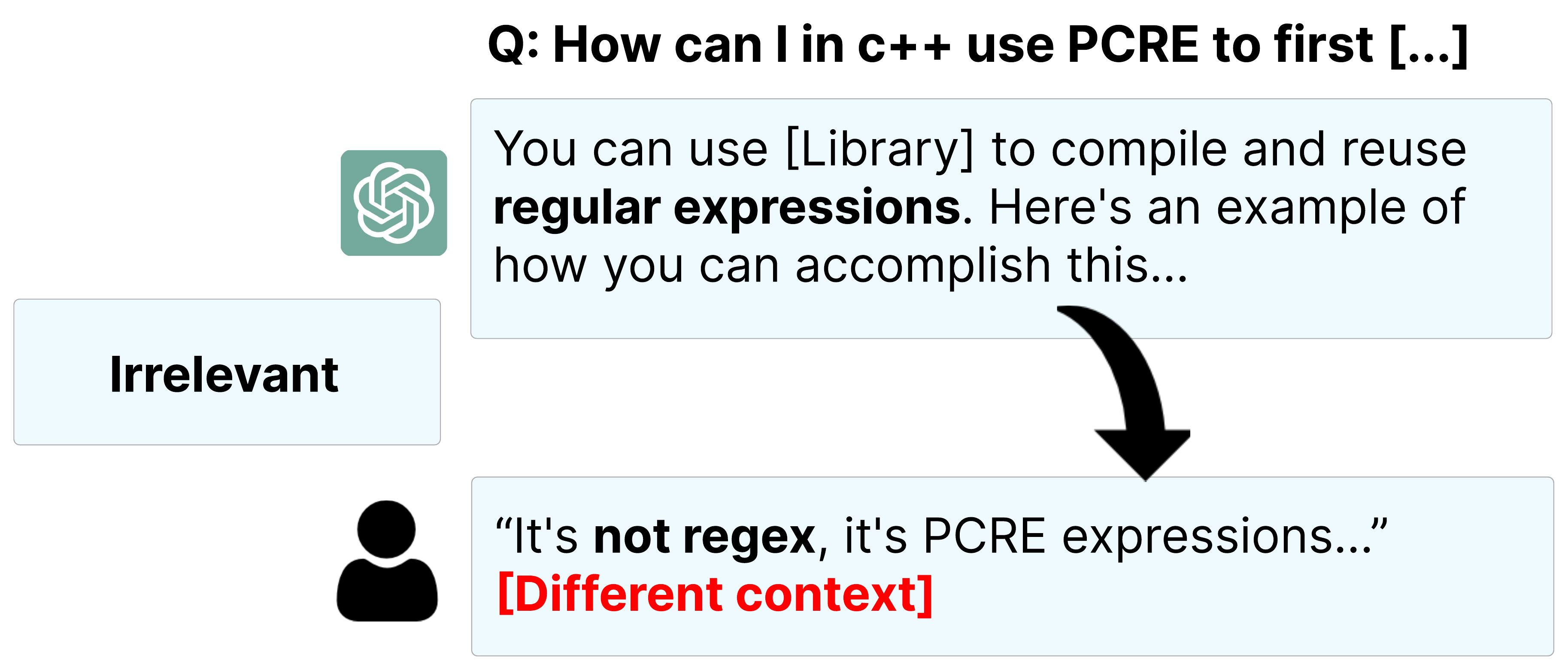}
    \caption{Example of Irrelevant Deficiencies in ChatGPT's Answers.}
    \label{fig:examples_irr}
\end{figure}

\textit{Irrelevant} answers are responses that were outdated or unrelated to the context of the issue (Figure \ref{fig:examples_irr}). These deficiencies were mostly seen in \textit{Tool/Library/API Recommendation} (n=8), \textit{Code Generation and Implementation} (n=6), and \textit{SE Information-Seeking} (n=4). In tool recommendation and code generation tasks, irrelevance is often related to the answer's inconsistency with the problem at hand. For example, a developer mentioned, \textit{``that is an entirely unrelated route for users to authenticate"}~\citep{irrelevant}. In information-seeking tasks, ChatGPT often provided outdated or unrelated information. For instance, a developer noted, \textit{``[Library] is mentioned in the documentation somewhere but noticed that is an archived repo so not maintained right?"}~\citep{irrelevant1}.

\begin{figure}[h]
    \centering
    \includegraphics[width=0.65\textwidth]{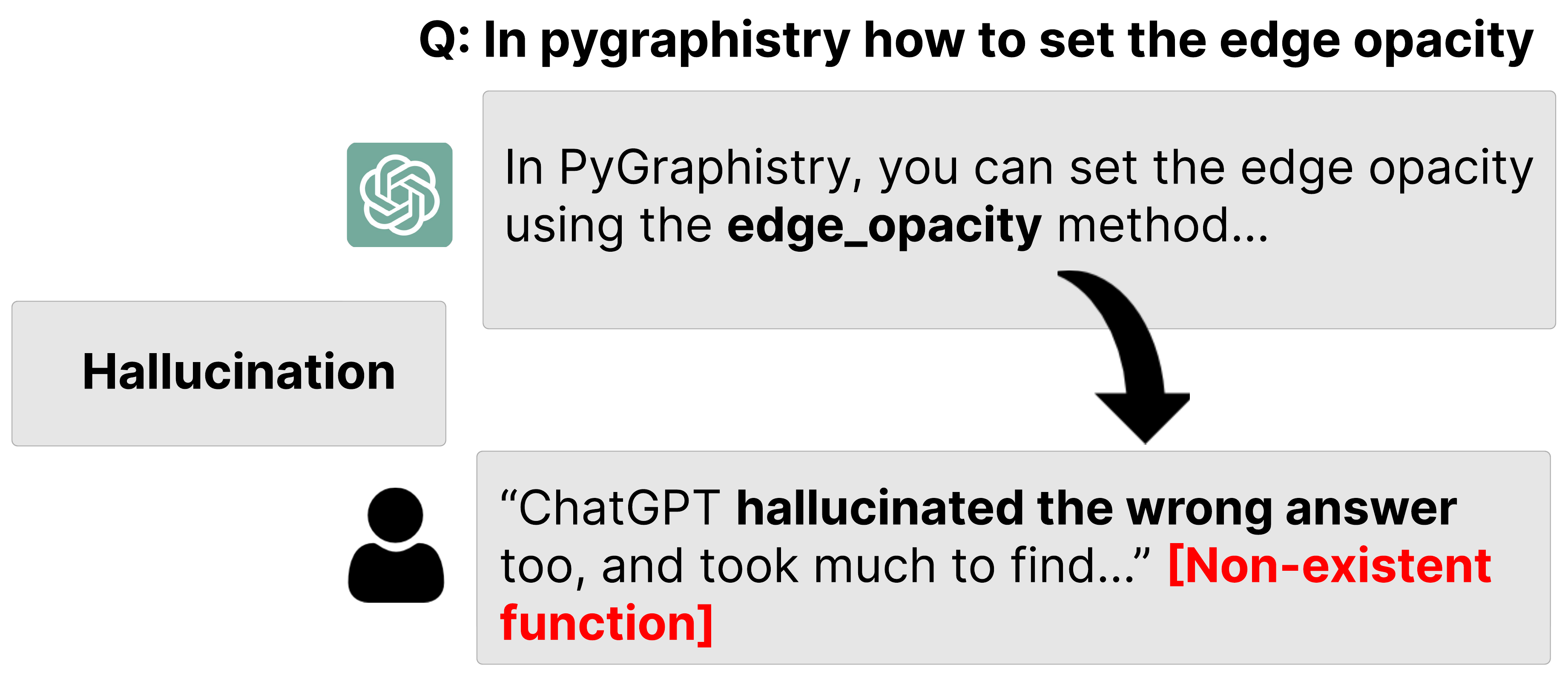}
    \caption{Example of Hallucination Deficiencies in ChatGPT's Answers.}
    \label{fig:examples_halluc}
\end{figure}

As for \textit{Hallucination}, the most frequent patterns involved the fabrication of non-existent methods, API calls within various libraries, and non-existent facts (Figure \ref{fig:examples_halluc}). Hallucinations were most commonly observed in \textit{Tool/Library/API Recommendation} (n=5) and \textit{SE Information-Seeking} (n=4). For example, a developer trying to use a library based on ChatGPT's suggestion noted, \textit{``ChatGPT gave a decent suggestion of what this might look like, but neither the source code nor the documentation has any hints that this exists"}~\citep{halluc}. Another instance was found in an issue thread where developers used ChatGPT as a search engine for their project, leading to a conclusion about the hallucinations in the answers: \textit{``...ChatGPT hallucinates, so is dead in the water as a reasonable search engine..."}~\citep{halluc1}.

\begin{tcolorbox}[]
The two most common deficiencies in ChatGPT's responses were information \textit{incorrectness} and lack of \textit{comprehensiveness}. 
Among other tasks, \textit{Tool/Library/API Recommendations} contained the highest number of deficiencies.
\end{tcolorbox}

\section{Discussion and Actionable Insights}
In this section, we present actionable insights for practitioners and researchers for issue resolution, focusing on:  a) interaction strategies with \textcolor{black}{conversational LLMs}, b) tools to optimize prompt design, and c) considerations for fine-tuning \textcolor{black}{conversational LLMs}. 

\noindent
\textbf{Interaction Strategies.} Our findings indicate that specific and concise prompts are key to effective issue resolution. Helpful conversations exhibit higher scores for contextual metrics (e.g., \textit{\#Code Snippets, \#Error Messages, \#URLs}) and readability metrics (e.g., \textit{SMOG Grade}), while avoiding verbosity with fewer \textit{\#Words} and \textit{\#Sentences}. These observations suggest that developers should prioritize providing detailed yet focused descriptions of issues, specifying key information, and avoiding broad or overly general queries.
One notable pattern in unhelpful conversations is the presence of lengthy prompts with multiple questions, large code snippets, and excessive or unfocused error messages. When addressing complex issues, developers can improve conversation outcomes by dividing the problem into discrete parts and tackling each individually. 

We suggest avoiding frequent topic changes within the same conversation, as our discourse metrics (e.g., \textit{Topic Coherence}) show that unhelpful interactions often shift between unrelated topics, leading to disjointed responses. 
In addition, providing feedback during the conversation enhances alignment with the developer's needs. \textcolor{black}{In helpful conversations, developers, especially those with more experience, often offered specific, targeted feedback on responses, indicating what was useful and highlighting areas for improvement. They were more effective at steering the conversation, clarifying ambiguous or incorrect responses, and breaking down complex tasks into smaller, manageable queries. This approach keeps \textcolor{black}{ChatGPT} focused and helps guide the conversation back on track if responses start to deviate.}

\noindent
\textbf{Tools for Prompt Optimization.} Our findings can inform 
development of tools to optimize \textcolor{black}{conversational LLM} prompts for effective issue resolution. Our conversational metrics—structural, linguistic, and discourse—can help flag and improve inefficient prompts  (e.g., automatically rephrasing issue descriptions to improve \textit{readability}, urging developers to \textit{include code snippets}). Tools providing real-time feedback and prompt optimization before developers submit queries to \textcolor{black}{ChatGPT} could significantly enhance response quality. 

\noindent
\textbf{Specialized \textcolor{black}{Conversational LLMs}.}
Our analysis shows that ChatGPT is more effective in tasks like \textit{Code Generation and Implementation}, \textit{Tool/Library /API Recommendation}, and \textit{Bug Identification and Fixing}, but struggles with \textit{Code Explanation} and \textit{SE Information-Seeking} related to issue resolution. Common issues, such as incorrect, irrelevant, incomplete, and unclear responses, were particularly prevalent in the \textit{Tool/Library/API Recommendation} and \textit{SE Information-Seeking} tasks. Our analysis also revealed instances of code-breaking suggestions and poor coding practices in ChatGPT responses, which further highlights the risks of using general-purpose \textcolor{black}{conversational LLMs} for SE-specific tasks. These findings highlight areas where ChatGPT performs well and where it fails to meet expectations. With these insights, we can focus on fine-tuning \textcolor{black}{conversational LLMs} to improve their performance in specific tasks, with the goal of developing specialized \textcolor{black}{conversational LLMs} for issue resolution.  

\noindent
\textcolor{black}{\textbf{Retrieval-Augmented Generation and Agentic Workflows.}
Many unhelpful conversations stemmed from a lack of project-specific knowledge or incomplete context. Our findings highlight how this gap strongly influences conversation outcomes, as reflected in contextual metrics such as the inclusion of \textit{\#Code Snippets}, \textit{\#Error Messages}, and \textit{\#URLs}. We also found that ChatGPT was less effective on issue types such as \textit{Debugging and Testing Issues} and \textit{Performance Optimization and Code Refactoring Issues}, which often require a deeper understanding of project internals, infrastructure, or system configurations. In such cases, Retrieval-Augmented Generation (RAG) can be used to ground ChatGPT’s responses in relevant documentation, codebases, or project artifacts. Automatically integrating retrieval pipelines into developer tools could enable ChatGPT to produce more context-aware responses without the need for full fine-tuning. By grounding responses in external sources of truth, RAG also helps reduce common deficiencies we identified, such as hallucinated content, factual inaccuracies, and incomplete or unclear suggestions~\citep{shuster-etal-2021-retrieval-augmentation, ayala-bechard-2024-reducing}.}
\textcolor{black}{In addition, issue resolution often requires handling multiple interconnected steps, something conversational LLMs struggle with in a single exchange. Our findings on issue difficulty show that ChatGPT is less effective on complex, time-consuming issues involving higher collaboration and longer resolution times. Agentic workflows offer a promising alternative by allowing the model to break down complex tasks into smaller subtasks, solve them step by step, and reflect on intermediate results along the way. Future systems that incorporate these agentic patterns may improve reliability and effectiveness for tasks such as test generation and multi-stage debugging. Together, RAG and agentic workflows represent practical and scalable ways to extend the capabilities of general-purpose conversational LLMs for real-world software engineering scenarios. }

%% file: 5_threats.tex
\section{Threats to Validity}

\noindent
\textbf{Construct Validity.} 
As with any study involving manual analysis, the results of this study are subjective to human judgment. To limit this threat, we ensured that the authors who analyzed the data had considerable experience in programming (3+ years) and qualitative analysis, and followed a consistent coding procedure that was piloted in advance. We conducted multiple rounds of coding and discussions to ensure consistency and resolve disagreements. Our average Cohen’s Kappa agreement was 0.88, indicating sufficient reliability.

\noindent
\textbf{Internal Validity.} The metrics that we used to measure conversational and project differences may have inherent biases or inaccuracies. For instance, 
how easy a conversation is to read i.e. \textit{readability} may differ from one developer to another, and 
the \textit{size of the project} might not directly affect the effectiveness of the usage of ChatGPT. To mitigate these threats, we used the Mann-Whitney test to statistically validate the differences between helpful and unhelpful conversations. 
By having multiple metrics for each category, conversational and project, we also ensured that our results were not biased by a single metric. Other potential threats include evaluation bias or script errors. To mitigate these, we performed sanity tests for all automated metrics.
\textcolor{black}{Additionally, while our multi-indicator approach to labeling helpfulness (including clone detection and manual inspection) helps reduce misclassification, there may still be rare edge cases where developer intent is ambiguous or misleading (e.g., expressions of gratitude despite unhelpful responses).}

\noindent
\textbf{External Validity.} Our findings are based on a dataset where developers shared links to their ChatGPT conversations within GitHub issue threads. Therefore, our results may not generalize to other ChatGPT conversations that developers choose not to share on GitHub. Our observations also might not generalize to developer conversations with other LLMs such as Llama and Copilot.
To mitigate these threats, we selected ChatGPT due to its widespread popularity and extensive use among developers.
In addition, our extensive dataset, collected from 655 projects comprising 48 programming languages, includes a diverse range of developer-ChatGPT interactions.
To further assess coverage, we conducted a follow-up crawl for the subsequent 13-month period (May 2024 – May 2025) since the start of our study. We found only 125 additional conversations, compared to the 876 in our original dataset. With 876 out of a known 1,001 total conversations, our sample achieves a ±1.2\% margin of error at the 95\% confidence level (assuming maximum variability), indicating that our dataset remains highly representative of the available population during the extended timeframe~\citep{cochran1977sampling}.

%% file: 2_background.tex
\section{Related Work}

\textbf{LLMs for Bug and Issue Resolution. }
LLMs, especially conversational LLMs, are increasingly used in bug localization, bug resolution, and issue tracking~\citep{hou2024large, wu2023large,tang2023empirical, ehsani2025hierarchicalknowledgeinjectionimproving}. Previous studies have investigated the effectiveness of LLMs in this context and found that developers often interact with ChatGPT to describe bug symptoms, provide contextual information, and seek guidance on potential solutions~\citep{das2024investigating}. Tao et al.~\citep{tao2024magisllmbasedmultiagentframework} developed an LLM-powered multi-agent framework to solve GitHub issue threads, achieving a resolution rate of 13.94\%.
Q\&A platforms like Stack Overflow have been crucial for assisting developers in issue resolution, but their traffic declined with the rise of conversational LLMs~\citep{dasilva2024chatgpt}. 
Several studies compared Q\&A forum answers to LLM responses, and found that LLMs struggle with questions related to unpopular frameworks and libraries~\citep{dasilva2024chatgpt}. For security questions, ChatGPT responses were correct only 56\% of the time~\citep{10260753}. Despite these concerns, users prefer ChatGPT for their well-articulated language~\citep{kabir2024stack}. ChatGPT answers were consistently found to be of lower quality than Stack Overflow, often lacking relevance and information timeliness~\citep{10298467}. Human expertise to solve specific exceptions makes Stack Overflow better in debugging tasks~\citep{liu2023better}.
Overall, these studies highlight the evolving role of LLMs in issue resolution. 
We contribute to the existing knowledge by identifying specific SE tasks that LLMs effectively support for issue resolution and evaluating their helpfulness for each task.

\textbf{Analysis of Developer-LLM Conversations. }
Understanding how software developers interact with ChatGPT has been a focus in recent works. 
The most common inquiries to ChatGPT are related to code generation, and obtaining answers to conceptual and how-to questions~\citep{hao2024empirical}, and the most commonly identified topics posed by developers are advanced programming guidance, information-seeking about frameworks, and obtaining high-level design recommendations \citep{MohamedChatting, SagdicDiscussion}. Developers often engage in multi-turn conversations to improve response quality by asking follow-up questions or refining prompts \citep{hao2024empirical}, LLM-generated code is primarily used for demonstrating high-level concepts or providing examples in documentation. Most of the conversations are about requesting improvements and more description inside generated codes \citep{jin_can_2024}. Mondal et al. identified 11 gaps that make ChatGPT conversations lengthy, with missing specifications and requests for additional functionality being the most frequent reasons \citep{mondal2024enhancing}. Champa et al. \citep{ChampaGHinAction} found that developers mostly seek assistance from ChatGPT with Python code for quality management and issue resolution tasks, and developer-ChatGPT conversations are the most efficient for software development management, optimization, and new feature implementation \citep{ChampaGHinAction}. Recent work has further revealed that developers primarily use ChatGPT for ideation, while direct reliance on its generated code to resolve issues is minimal. The study also identified areas like refactoring and data analytics where ChatGPT excels, and debugging and automation tasks where dissatisfaction is higher \citep{das2024developersengagechatgptissuetracker}.
Our paper complements existing work by providing insight into the dynamics and characteristics that make developer-ChatGPT conversations successful in the context of issue resolution.

%% file: 6_conclusion.tex
\section{Conclusion and Future Work}
In this study, we explored the usage and effectiveness of conversational LLMs for issue resolution within GitHub's issue threads. 
We found that only 62\% of the developer-ChatGPT conversations helped in resolving issues across tasks such as \textit{Code Generation and Implementation}, \textit{Tool/Library/API Recommendation}, and \textit{Bug Identification and Fixing}.
We identified that helpful and unhelpful conversations have key conversational differences in their structure, linguistic patterns, and discourse. Project metrics show that larger and more popular projects, as well as more experienced developers, use ChatGPT more effectively for issue resolution. \textcolor{black}{Our issue-level analysis further reveals that ChatGPT is more effective on tasks such as feature requests and compilation errors, but tends to struggle on complex, time-consuming problems involving multiple developers, such as debugging and project refactoring.} Our findings provide actionable insights for practitioners and researchers, including strategies for effective interactions, guidance for developing tools to optimize prompts, and recommendations for fine-tuning LLMs for issue resolution tasks.

In the future, we plan to experiment with fine-tuned LLMs and develop a system that streamlines interactions with conversational LLMs by integrating identified metrics into developer prompts. 
We will conduct experiments to evaluate whether these enhanced interactions lead to better issue resolution outcomes and a more positive user experience.
\\
\\
\noindent
\textbf{Data Availability: }The replication package of our study, including the datasets, code, and analysis instructions are available at: \url{https://figshare.com/s/2a7284f4c593888a35cc}
\\
\\
\textbf{Competing Interests: }The authors declare that they have no conflict of interest.
\\
\\
\textbf{Funding: }No funding was received for conducting this study.
\\
\\
\textbf{Ethics Declaration: }Not applicable.